\documentclass[11pt,a4paper]{article}
\usepackage[utf8]{inputenc}
\usepackage{adjustbox}
\usepackage[margin=1in]{geometry}
\usepackage{amsmath,amssymb,amsthm}
\usepackage{graphicx,xcolor,colortbl}
\usepackage{paralist,booktabs,tabularx}
\usepackage[basic]{complexity}
\usepackage{todonotes,xspace,wasysym}
\usepackage{cite,float,microtype,fmtcount}
\usepackage[shortlabels]{enumitem}
\usepackage[font=small,labelfont=small]{subcaption}
\usepackage[font=small,labelfont=bf,format=plain]{caption}
\usepackage[pdfpagelabels,colorlinks,allcolors=blue]{hyperref}

\title{External Labeling Techniques: A Taxonomy and Survey}

\author{Michael A. Bekos$^{1}$, Benjamin Niedermann$^{2}$, Martin N\"ollenburg$^{3}$
\\[0.1in]
$^1$\small Institute for Informatics, University of T{\"u}bingen, T{\"u}bingen, Germany\\
\texttt{\small bekos@informatik.uni-tuebingen.de}
\and
$^2$\small Institute of Geodesy and Geoinformation, University of Bonn, Bonn, Germany\\
\texttt{\small niedermann@igg.uni-bonn.de}
\and
$^3$\small Algorithms and Complexity Group, TU Wien, Vienna, Austria\\
\texttt{\small noellenburg@ac.tuwien.ac.at}
}
\date{}

\newcommand{\ls}{\texttt{s}}

\newcommand{\lo}{\texttt{o}}
\newcommand{\lp}{\texttt{p}}
\newcommand{\ld}{\texttt{d}}
\newcommand{\lr}{\texttt{r}}

\newcommand{\lpo}{\lp\lo}
\newcommand{\ldo}{\ld\lo}
\newcommand{\lod}{\lo\ld}
\newcommand{\lpd}{\lp\ld}
\newcommand{\lopo}{\lo\lp\lo}

\newcommand{\excentricLabeling}{\ocircle}

\newcommand{\OPT}{\mathrm{OPT}}
\newcommand{\rcite}[1]{\rotatebox{90}{\cite{#1}}}
\newcommand{\ocite}[1]{\rotatebox{90}{\cite{#1}$\ocircle$}}
\newtheorem{problem}{Problem}
\newtheorem{op}{Challenge}

\newcommand{\slAll}{29\xspace}
\newcounter{slAllCounter}
\newcommand{\slRectangleLabel}{S1.1\xspace}

\newcommand{\slCircleLabel}{S1.2\xspace}

\newcommand{\slConvexHullLabel}{S1.3\xspace}

\newcommand{\slOtherHullLabel}{S1.4\xspace}

\newcommand{\slIntersectionLabel}{S3.1\xspace}
\newcommand{\slIntersection}{18\xspace}

\newcommand{\slIntersectionNotI}{11}
\newcommand{\slLengthLabel}{S3.2\xspace}

\newcommand{\slLengthNotI}{2}
\newcommand{\slSpacingLabel}{S3.3\xspace}

\newcommand{\slDirectionLabel}{S3.4\xspace}

\newcommand{\slDirectionI}{15}

\newcommand{\slAlignmentLabel}{S3.5\xspace}

\newcommand{\slGroupingLabel}{S3.6\xspace}

\newcommand{\slObstaclesLabel}{S3.7\xspace}

\newcommand{\slMixedLabel}{S3.8\xspace}

\newcommand{\slMixedI}{3}

\newcommand{\slSchemeLabel}{S5.1\xspace}
\newcommand{\slScheme}{14\xspace}

\newcommand{\slGreedyLabel}{S5.2\xspace}

\newcommand{\slGreedyI}{12}

\newcommand{\slForceBasedLabel}{S5.3\xspace}

\newcommand{\slForceBasedI}{8}

\newcommand{\slDPLabel}{S5.4\xspace}

\newcommand{\slMetaHeuristicLabel}{S5.5\xspace}

\newcommand{\slMatchingLabel}{S5.6\xspace}

\newcommand{\slSweepLabel}{S5.7\xspace}

\newcommand{\slImpLabel}{S6.1\xspace}
\newcommand{\slImp}{27\xspace}

\newcommand{\slMALabel}{S6.2\xspace}

\newcommand{\slMAI}{6}

\newcommand{\slProofsLabel}{S6.3\xspace}

\newcommand{\slProofsI}{6}

\newcommand{\slEILabel}{S6.4\xspace}

\newcommand{\slEII}{4}

\newcommand{\slUSLabel}{S6.5\xspace}

\newcommand{\slUSI}{6}

\newcommand{\slHeuristic}{24\xspace}

\newcommand{\slExactI}{8}

\newcommand{\slCommAlgoI}{3}

\newcommand{\slCommVis}{26\xspace}

\newcommand{\polyAll}{25\xspace}
\newcounter{polyAllCounter}
\newcommand{\plPOLabel}{P1.1\xspace}

\newcommand{\plOPOLabel}{P1.2\xspace}

\newcommand{\plDOPDLabel}{P1.3\xspace}

\newcommand{\plOtherSlopeLabel}{P1.4\xspace}

\newcommand{\plLengthLabel}{P2.1\xspace}

\newcommand{\plBendsLabel}{P2.2\xspace}

\newcommand{\plMultiLabel}{P2.3\xspace}

\newcommand{\plOtherLabel}{P2.4\xspace}

\newcommand{\plOneLabel}{P3.1\xspace}

\newcommand{\plTwoLabel}{P3.2\xspace}
\newcommand{\plTwo}{14\xspace}

\newcommand{\plTwoAdjLabel}{P3.3\xspace}

\newcommand{\plThreeLabel}{P3.4\xspace}

\newcommand{\plFourLabel}{P3.5\xspace}

\newcommand{\plotherContourLabel}{P3.6\xspace}

\newcommand{\plUniformLabel}{P4.1\xspace}
\newcommand{\plUniform}{21\xspace}

\newcommand{\plNonuniformLabel}{P4.2\xspace}
\newcommand{\plNonuniform}{12\xspace}

\newcommand{\plFixedposLabel}{P5.1\xspace}

\newcommand{\plSlidingposLabel}{P5.2\xspace}

\newcommand{\plFixedportLabel}{P6.1\xspace}

\newcommand{\plSlidingportLabel}{P6.2\xspace}

\newcommand{\plProofsLabel}{P8.1\xspace}

\newcommand{\plImpLabel}{P8.2\xspace}

\newcommand{\plUserStudyLabel}{P8.3\xspace}

\newcommand{\plExactLabel}{P9.1\xspace}
\newcommand{\plExact}{21\xspace}

\newcommand{\plApproxLabel}{P9.2\xspace}

\newcommand{\plHeuristicLabel}{P9.3\xspace}

\newcommand{\plDPLabel}{P10.1\xspace}

\newcommand{\plSweepLabel}{P10.2\xspace}

\newcommand{\plMatchingLabel}{P10.3\xspace}

\newcommand{\plSchedLabel}{P10.4\xspace}

\newcommand{\plLocalLabel}{P10.5\xspace}

\newcommand{\plILPLabel}{P10.6\xspace}

\newcommand{\plMetaLabel}{P10.7\xspace}

\newcommand{\plGreedyLabel}{P10.8\xspace}

\newcommand{\plOtherTechLabel}{P10.9\xspace}

\newcommand{\plComplexityLabel}{P10.10\xspace}

\newcommand{\plAlgLabel}{P11.1\xspace}
\newcommand{\plAlg}{17\xspace}

\newcommand{\plVisLabel}{P11.2\xspace}
\newcommand{\plVis}{8\xspace}

\newcounter{allReferences}
\addtocounter{allReferences}{\value{slAllCounter}}
\addtocounter{allReferences}{\value{polyAllCounter}}

\newcommand{\transnum}[1]{\ifnum#1>12 {#1} \else {\numberstringnum{#1}}\fi}
\newcommand{\transNum}[1]{\ifnum#1>12 {#1} \else {\Numberstringnum{#1}}\fi}

\newcolumntype{a}{>{\centering\arraybackslash}p{0.3ex}}
\newcolumntype{b}{>{\columncolor{lightgray}\centering\arraybackslash}p{0.3ex}}

\let\tempone\itemize
\let\temptwo\enditemize
\renewenvironment{itemize}{\tempone\addtolength{\itemsep}{-0.5\baselineskip}}{\temptwo}

\newcommand{\publications}[1]{{\medskip\noindent \emph{Publications:} #1.}}

\begin{document} 

\maketitle

\begin{abstract}
External labeling is frequently used for annotating features in graphical displays and visualizations, such as technical illustrations, anatomical drawings, or maps, with textual information. Such a labeling connects features within an illustration by thin leader lines with their labels, which are placed in the empty space surrounding the image. Over the last twenty years, a large body of literature in diverse areas of computer science has been published that investigates many different aspects, models, and algorithms for automatically placing external labels for a given set of features. This state-of-the-art report introduces a first unified taxonomy for categorizing the different results in the literature and then presents a comprehensive survey of the state of the art, a sketch of the most relevant algorithmic techniques for external labeling algorithms, as well as a list of open research challenges in this multidisciplinary research field.
\end{abstract}  

\section{Introduction}

\begin{figure*}[t]
\centering
\begin{subfigure}[b]{0.24\linewidth}
  \centering \includegraphics[trim = 0mm 25mm 0mm
  25mm,clip,width=.8\linewidth,page=1]{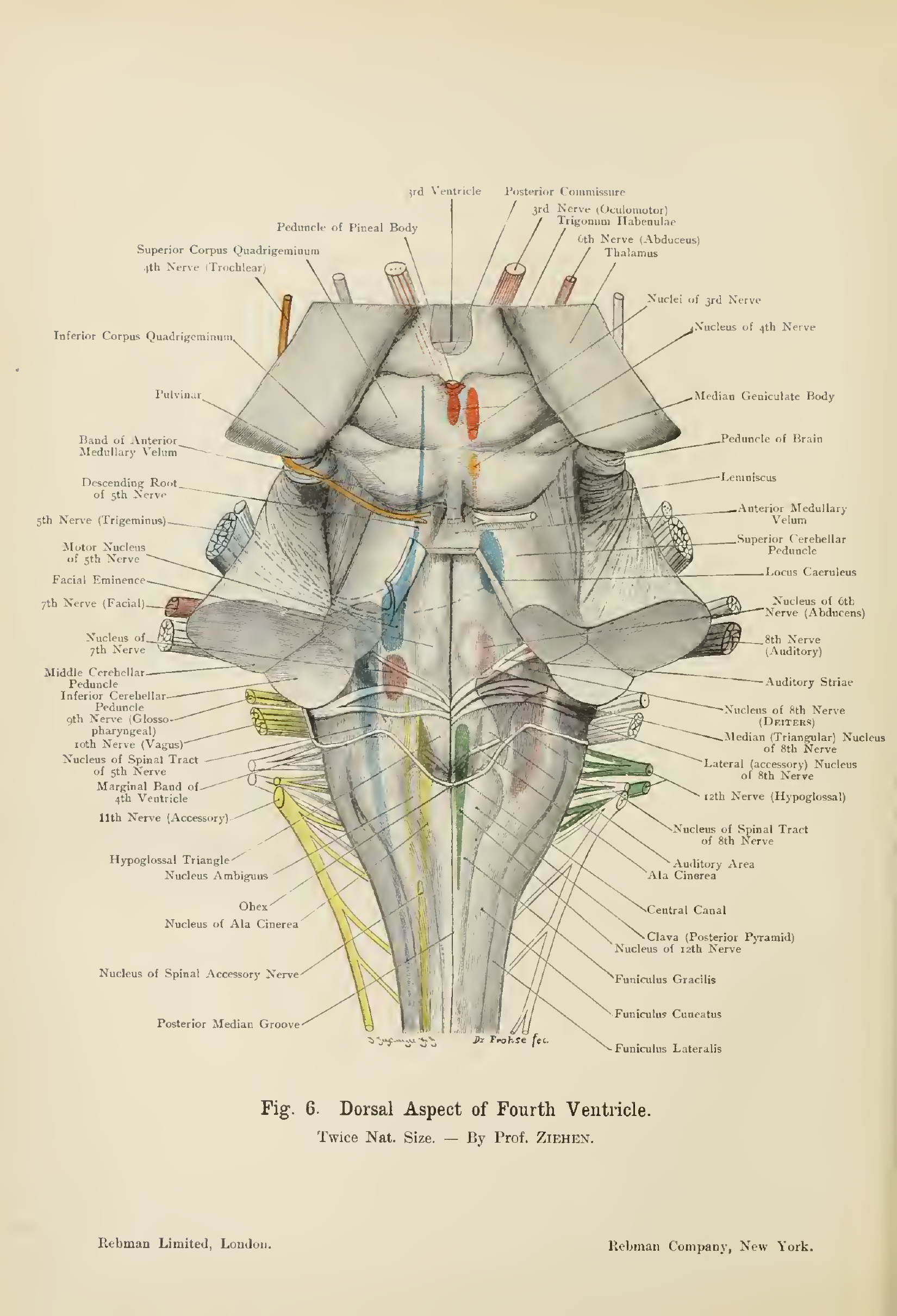}
  \subcaption{\label{fig:external-labeling}}
  \end{subfigure}
  \hfill
  \begin{subfigure}[b]{0.24\linewidth}
    \centering
    \includegraphics[page=8]{fig/layer-decomposition}
    \subcaption{\label{fig:positions:contour-labeling}}
  \end{subfigure}
\hfill
  \begin{subfigure}[b]{0.24\linewidth}
    \centering
    \includegraphics[page=9]{fig/layer-decomposition}
    \subcaption{\label{fig:positions:boundary-labeling}}
  \end{subfigure}
  \hfill
  \begin{subfigure}[b]{0.24\linewidth}
    \centering
    \includegraphics[page=10]{fig/layer-decomposition}
    \subcaption{\label{fig:positions:excentric-labeling}}
  \end{subfigure}  
  \caption{External labelings. (\subref{fig:external-labeling})~Anatomical illustration from: \emph{Atlas of applied
      (topographical) human anatomy for students and
      practitioners.} 
    \emph{K. H. v. Bardeleben, 
      E. Haeckel}, 
      Rebman Company, 1906. 
    (\subref{fig:positions:contour-labeling})~Contour
    labeling: the positions of the reference points of the leaders are restricted to a
    predefined contour
    (dashed). (\subref{fig:positions:boundary-labeling})~Boundary
    labeling: the positions of the reference points of the leaders are restricted to a
    rectangle.  (\subref{fig:positions:excentric-labeling})~Excentric
    labeling: the labeled features are contained in a circle (blue), while
    the labels are placed around the circle.
  }\label{fig:examples}
\end{figure*}

Displaying textual annotations (or \emph{labels}) for graphical features  in maps, technical illustrations like assembly instructions or cutaway illustrations, or anatomy drawings is an important part of information visualization. One can broadly distinguish \emph{internal} and \emph{external} labeling, where internal labels are placed inside or in the direct neighborhood of a feature, whereas external labels are placed in the margins outside the illustration, where they do not occlude the illustration itself; see Fig.~\ref{fig:external-labeling} for an example of external labeling used in anatomy drawings. In order to link each label to its respective feature, thin leader curves are used that connect label-feature pairs in the form of  \emph{callouts}; see~Fig.~\ref{fig:examples}.

Over the last twenty years various aspects of the external labeling problem have been considered in computer science -- both from a theoretical and from a practical point of view. The range of topics is so wide that the term ``external labeling'' mostly serves as an umbrella term, which covers several different labeling approaches, such as \emph{contour labeling}~\cite{Ali2005,Niedermann2017} (see Fig.~\ref{fig:positions:contour-labeling}), \emph{boundary labeling}~\cite{Bekos2007} (see Fig.~\ref{fig:positions:boundary-labeling}), and \emph{excentric} or \emph{focus-region labeling}~\cite{Fekete1999,Bruckner2005} (see Fig.~\ref{fig:positions:excentric-labeling}). Preliminary surveys on external labeling take a rather narrow view on selected labeling models, and focus either on selected theoretical results~\cite{k-lwl-09,w-gdc-13,kt-aplmd-18}, or on practical works~\cite{Oeltze-Jafra2014}. 

In this state-of-the-art report, we introduce a unified and extensible
taxonomy for the different parameters and models found in the external
labeling literature (Section~\ref{sec:model}). Even though the focus
of this report is on the algorithmic techniques for external labeling,
Section~\ref{sec:visual-aspects} first gives a summary of the most
important visual aspects to be considered in external labeling. Then,
based on the new taxonomy, we present a comprehensive literature
survey of more than 50 research papers\footnote{On
  \url{www.externallabeling.net} we maintain a visual bibliography on external labeling techniques.} on external labeling algorithms
(Section~\ref{sec:star}), ranging from theoretical and algorithmic
contributions to practical research and results in visual computing,
including some empirical studies. From this
literature, we collected a set of the most frequently applied
algorithmic techniques, from exact, geometric algorithms to flexible,
heuristic approaches, that are well suited to address external
labeling problems. We provide sketches of the main techniques
(Section~\ref{sec:techniques}) and discuss their properties and
limitations as a guideline for researchers and practitioners
interested in external labeling.  We conclude with a discussion of ten
future challenges in external labeling from various technical
perspectives (Section~\ref{sec:discussion}).

The purpose of our state-of-the-art report is to provide a broad overview on this diverse and multi-disciplinary research topic, both for researchers entering the area or those extending their scope and for practitioners and domain users searching for suitable labeling techniques for their particular data visualization tasks. Our taxonomy will help to give structure to existing and future research results in external labeling, where previous attempts have been too narrow (e.g., only addressing boundary labeling) or even missing completely. 
The detailed literature survey gives a systematic overview of the existing knowledge collected in more than 50 publications. 
The high-level discussion of the most used algorithmic techniques in external labeling serves as an orientation aid when deciding which existing algorithm to select and implement or when designing a new labeling algorithm for a particular labeling task.
Typically generic techniques are simpler to adapt to new settings and faster to implement, whereas specialized algorithms show better performance but are more difficult to adapt and implement.
Finally, our collection of open research challenges provides ample opportunities for future work on external labeling. The challenges range from algorithmic problems and less explored labeling models, e.g., combining external and internal labeling, to perceptual questions and studies of human comprehensibility of the labeled data.

\begin{figure*}[t]
  \centering
  \includegraphics[width=\textwidth]{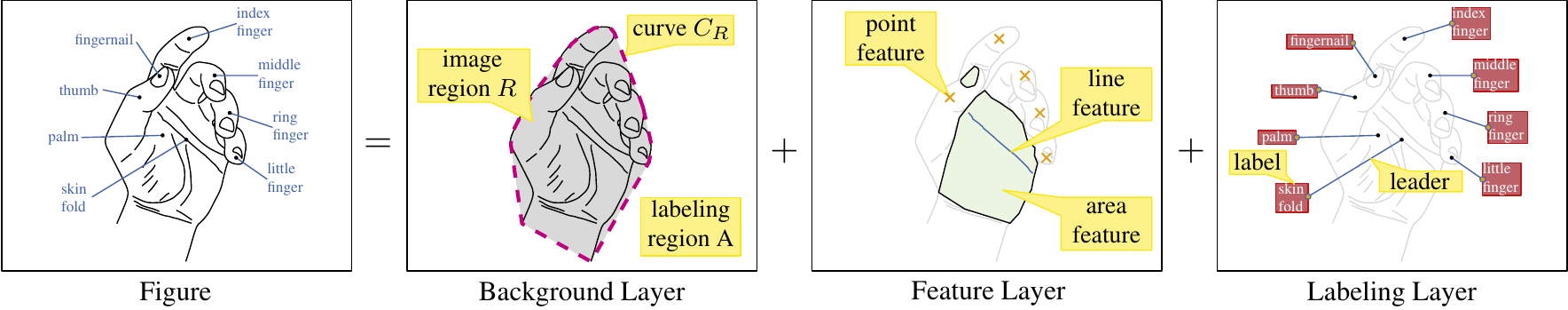} 
  \caption{Layer decomposition. The figure (left) is decomposed into the background, feature, and labeling layer.}
  \label{fig:layer-decomposition}
\end{figure*}

\section{A unified taxonomy}\label{sec:model}
This section introduces a unified and extensible taxonomy of different labeling
models proposed in the literature and may serve as an entry point for
the interested reader who wants to get familiar with the basic
concepts of external labeling.  In
contrast to previous, narrower taxonomies (e.g., on boundary labeling~\cite{BekosKPS06}), we
provide a common, general structure for classifying existing and future results in the wider field of external labeling. It is designed to be general and extendable enough for addressing new challenges
in external labeling and covers all aspects of existing work. 
First, we formally introduce the most important terminology and
concepts on external labeling
(Section~\ref{sec:model:basic}). Afterwards, we present distinctive
features that we use to characterize and unify the different models
for external labeling found in the literature
(Section~\ref{sec:model:features}). Finally, we formally define a
generic optimization problem that underlies approaches for creating
external labelings automatically
(Section~\ref{sec:model:optimization}).

\subsection{Terminology and concepts}\label{sec:model:basic}

In general a figure with external labeling can be decomposed into
three overlaid layers; see Fig.~\ref{fig:layer-decomposition}. The bottom
layer defines the image area containing the illustration and the
labeling area. The middle layer contains the features to be
labeled. The top layer finally contains the annotations for the
features.  

\paragraph*{Background layer} The bottom layer partitions the available
\emph{drawing area} $D$ along a simple, closed curve~$C_R$ into an \emph{image
region} $R$ and a \emph{labeling region} $A=D\setminus R$; we call this layer
the \emph{background layer}. The image region contains the unlabeled
illustration, while the labeling region defines the empty space
surrounding the image, which is reserved for placing the
annotations. For instance,~the image region could be a bounding box or
a convex hull of the illustration.

\paragraph*{Feature layer} The middle layer defines a set $F$ of
pairwise disjoint features of interest to be labeled; we call this layer the
\emph{feature layer}. Each feature is a small sub-region of $R$, which
can be a point, a curve or an area. Further, each feature has some attached
information, e.g., its name, an icon or a short textual description,
which will later be used as the content of its label. Geometrically,
we define the \emph{label} as the axis-aligned bounding box of the
feature's attached information. We denote the set of all labels by
$L$. Typically, each feature has its own label,~but in some special
cases multiple features can share the same label or one feature can have
several~labels.

\begin{figure}[t]
  \centering
  \includegraphics[page=3]{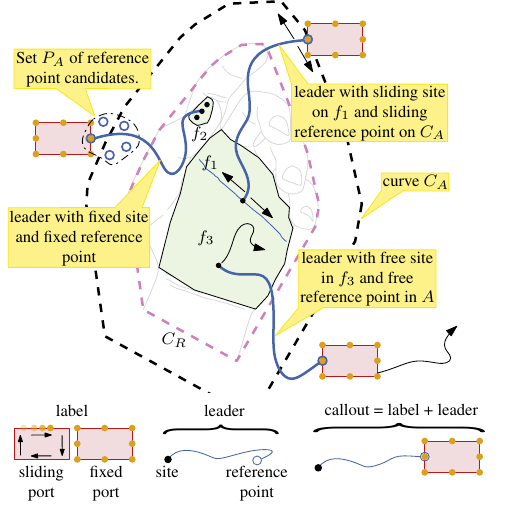} 
  \caption{Terminology used throughout this paper.}
  \label{fig:labeling-concepts}
\end{figure}

\paragraph*{Labeling layer} The top layer finally adds the labels and
leaders to the figure; we call this layer the \emph{labeling layer},
refer also to Fig.~\ref{fig:labeling-concepts} for a detailed example.  More
precisely, the label~$\ell$ of a feature~$f$ is placed within the
labeling region $A$ and connected to $f$ by a simple curve~$\lambda$;
we call $\lambda$ the \emph{leader} of $\ell$ and $f$. One end point
of~$\lambda$, the so-called \emph{site} (sometimes also called \emph{anchor}) $\sigma(\lambda)$, is a point
contained in $f$. 
The other end point of $\lambda$, the so-called \emph{reference point}
$a(\lambda)$, is a point on the boundary of $\ell$. We distinguish
different ways of placing sites, reference points of leaders and
labels. If the site~$\sigma$ can be any point of $f$, then $\sigma$ is
a \emph{free} site. If $\sigma$ is restricted to be any point of a
pre-defined curve in $f$, then $\sigma$ is a \emph{sliding}
site. Finally, if $\sigma$ is restricted to be any point of a finite
set of points in $f$, then $\sigma$ is a \emph{fixed} site. Note that
the site of a point feature is always fixed, in fact the site and the
feature coincide. Similarly, if the reference point $a$ can be any
point in $A$, then $a$ is a \emph{free} reference point. If $a$ is
restricted to be any point of a curve~$C_A$ in $A$, then $a$ is a
\emph{sliding} reference point.  Finally, if $a$ is restricted to be
any point of a finite set~$P_A$ of points in $A$, then $a$ is a
\emph{fixed} reference point. The most common case is the use of
sliding or fixed reference points, where the curve~$C_A$ or the set
$P_A$ either coincide with $C_R$ or are offset by a small distance
from $C_R$. The case of free reference points is less relevant,
because labels should generally be placed close to the image region.
Recall that the reference point $a$ is a point on the boundary of the
label~$\ell$. From the perspective of~$\ell$ we call this
attachment point between the leader $\lambda$ and $\ell$ the
\emph{port}~$\pi(\ell)$ of~$\ell$, which implies $a(\lambda)=\pi(\ell)$.
If the port $\pi(\ell)$ can be any point on the boundary of $\ell$,
then $\pi(\ell)$ is a \emph{sliding} port. If $\pi(\ell)$ is
restricted to be any point of a finite set of points on the boundary
of $\ell$, then $\pi(\ell)$ is a \emph{fixed} port. The most common
fixed-port model restricts the ports to the corners or midpoints of
the edges of $\ell$. If also the label positions are fixed, 
the reference point of $\lambda$ is limited to the ports of $\ell$. 
We call the geometric composition
  $\gamma=(\lambda,\ell,\pi(\ell))$ of a leader $\lambda$ and a placed
  label $\ell$ attached to each other at the port $\pi(\ell)$ a \emph{callout}. A
set~$\mathcal L$ of callouts is called an \emph{external labeling} of
the feature set~$F$; for brevity we also call it \emph{labeling}. We
say that $\mathcal L$ is \emph{plane} or \emph{crossing-free} if no
two callouts in $\mathcal L$ intersect.

\subsection{Distinctive features}\label{sec:model:features}
In this section we discuss three distinctive features of external
labeling that are used to characterize the different labeling models
found in the external labeling literature.

\paragraph*{Admissible positions of labels}
The admissible positions of the reference points of the leaders and accordingly of the labels
significantly influence the overall appearance of the labeled figure.  While
in general models for external labeling the reference points may be placed
anywhere in the labeling area, models for \emph{contour labeling}
restrict the positions of the reference points to a predefined contour $C_A$;
see Fig.~\ref{fig:positions:contour-labeling}. This contour is often chosen in such a way that it
roughly matches the shape of the illustration in the background layer. This
is a common technique to achieve a label placement that blends in with
the illustration.

An even more restrictive model is that of \emph{boundary labeling}, which requires
that the image region~$R$ is a rectangle and $C_A$ coincides with the
boundary of $R$; see
Fig.~\ref{fig:positions:boundary-labeling}. Hence, excluding the
corners of $R$ for the placement of reference points, each label
intersects exactly one edge of the image boundary in more than one
point; we say that the label \emph{touches} that edge.  Depending on
the number of edges that are touched by labels, we distinguish
\emph{1-sided}, \emph{2-sided}, \emph{3-sided} and \emph{4-sided}
boundary labeling models. We observe that for 2-sided boundary
labeling models the edges that are touched by labels are either
\emph{adjacent} or \emph{opposite} edges of $R$;
Figure~\ref{fig:positions:boundary-labeling} shows a 2-sided boundary
labeling with opposite edges touched by labels.

In \emph{excentric labeling} only features in a restricted region of
the illustration are labeled with their labels placed
outside that region; see
Fig.~\ref{fig:positions:excentric-labeling}. In spite of labels possibly
overlapping the actual illustration in the background, we consider
excentric labeling to be a special case of external labeling, because
the labels are still clearly separated from their features. In excentric labeling the restricted region is typically described by a
circle of fixed radius and implements the metaphor of a \emph{lens}
that can be moved over the illustration to explore its
details. Alternatively, one can think of this restricted region as a
\emph{focus region} of the user. In the literature, two main variants
are found: labels that are placed freely in the surroundings of the
focus region and the special case of contour labeling requiring that
all labels are placed along the boundary of the focus region.

\paragraph*{Leader type} 
The model introduced in Section~\ref{sec:model:basic} allows
general curves for leaders without any restriction to their
shape. However, in actual externally labeled figures only leaders with
simple, schematic shapes, such as straight lines and polylines, are used
in order to avoid cluttered illustrations and to sustain
legibility. In rare cases, \emph{curved} leaders such as Bézier arcs are used. Next, we present a systematic classification of different polyline
leader types. 
Let~$\lambda$ be a
leader that consists of $k$ segments, which are ordered from its site
to its reference point. We describe the shape of $\lambda$ by a string
$z\in\{\ls,\lr,\ld,\lo,\lp\}^k$ of $k$ symbols, where the
$i$-th symbol of $z$ describes the orientation of the $i$-th segment
of $\lambda$. In the following we list the meaning of each symbol.
\begin{itemize}
\item[\textbf{\ls}] An $\ls$-segment is a straight-line segment with unrestricted slope.
\item[\textbf{\lr}] An $\lr$-segment lies on a ray that emanates from a
  given center point $M$.
\item[\textbf{\lo}] An $\lo$-segment is orthogonal to a reference line or line segment~$m$.
\item[\textbf{\lp}] A $\lp$-segment is parallel to a reference line or line segment~$m$.
\item[\textbf{\ld}] A $\ld$-segment~$l$ (for a given angle
  $0^\circ < \alpha < 90^\circ$) is a diagonal segment defined relative to its preceding
  segment $l_p$ and its succeeding segment $l_s$ as follows. The
  turning angle between $l_p$ and $l$ as well as the turning angle
  between $l$ and $l_s$ is $\pm \alpha$, where the turning angle of two segments sharing a common endpoint is the minimum angle in which one has to rotate one of the two segments around their common endpoint to reach the second.
\end{itemize}

\begin{figure}[t]
  \centering
  \begin{subfigure}[b]{0.49\linewidth}
    \centering
    \includegraphics[page=4]{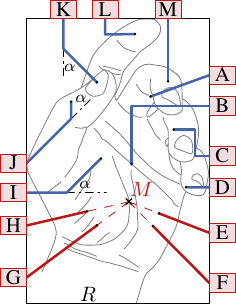}
    \subcaption{\label{fig:leader-types:overview}}
  \end{subfigure}
   \begin{minipage}[b]{0.47\linewidth}
     \begin{subfigure}[b]{\linewidth}
       \centering
       \includegraphics[page=7]{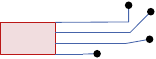}
       \subcaption{\label{fig:leader-types:hyper-leader}}
      \end{subfigure}
  
      \begin{subfigure}[b]{\linewidth}
       \centering
       \includegraphics[page=6]{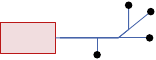}
       \subcaption{\label{fig:leader-types:hyper-leader-schematic}}
    \end{subfigure}
  \end{minipage}

  \caption{Types of
    leaders. (\subref{fig:leader-types:overview})~Leader types mostly
    used in
    literature.
    (\subref{fig:leader-types:hyper-leader}) A label connected to multiple sites. (\subref{fig:leader-types:hyper-leader-schematic})~A
    hyperleader, i.e., a set of leaders having the same reference point, but
    different sites. }
\end{figure}

Due to their shape, leaders consisting only of $\lo$- and
$\lp$-segments are called \emph{orthogonal} leaders, when
the reference line or line segment is either vertical or horizontal. 
If one additional type of
$\ld$-segments is allowed, then the resulting leaders are called
\emph{orthodiagonal}. In the special case, in which the turning angles
of each $\ld$-segment with its preceding and succeeding segments are
$\pm 45^\circ$, the obtained leaders are called \emph{octilinear}
following the established terminology in network visualization.
In models for boundary labeling the shape of a leader is typically
described with respect to the edge of the boundary rectangle
$R$ that is touched by the leader's label.
Following this convention, Figure~\ref{fig:leader-types:overview} illustrates the most frequent leader types in the literature. 
The labels A, E, F, G and H are connected to $\ls$-leaders. 
More precisely, while A is connected to an
arbitrary $\ls$-leader, the labels E, F, G, and H have
$\lr$-leaders, because their straight-line segments lie on rays
emanating from the center $M$.
Further, the labels D and M have $\lo$-leaders. 

The remaining labels B, C, I, J, K, and L have leaders with more than
one segment. Assuming again that the reference line segment of each leader is the side of the rectangle containing its label,  B has a $\lp\lo$-leader,  C has an
$\lo\lp\lo$-leader, and  L has an $\lp\lo$-leader; recall that
the segments of a leader are ordered from its site $\sigma$ to its reference point $a$. Further, label I has a $\ld\lo$-leader, label J has a
$\lp\ld$-leader, and label K has a $\ld\lo$-leader. In all three
cases the $\ld$-segment has a turning angle $\alpha$ with respect to
its preceding and succeeding segment, respectively; with rare exceptions $\alpha=45^\circ$ is used for $\ld$-segments in the literature.

In some applications multiple features can share the same label, i.e.,
a single label is connected to multiple features via multiple leaders;
see Fig.~\ref{fig:leader-types:hyper-leader}. Using this labeling
style the reader can easily identify features with the same
meaning. Further, it saves space in the labeling area. We refer to
this labeling technique as \emph{many-to-one labeling}. Further, by
connecting all leaders to the same reference point of the label,
the leaders can be bundled; we call the set~$\Lambda$ of these leaders
a \emph{hyperleader} of the label; see
Fig.~\ref{fig:leader-types:hyper-leader-schematic}. The literature on
labeling with hyperleaders requires that no two leaders in $\Lambda$
intersect apart from a common suffix. This leads to the desired visual
impression that $\Lambda$ is a single leader forking to multiple
sites.

\paragraph*{Temporal dimension}
A widely used application of external labeling are digital
visualizations that change over time. For example, consider a medical
visualization system that provides the interactive exploration of a
3D body model and that uses external labeling to explain
the features of that model. When the user changes the view of
the 3D model, the labeling changes correspondingly.  From a
technical point of view, this process can be seen as an animation
consisting of a temporal sequence of 2D images that arise
from projecting the 3D model onto the screen space. For each
of these 2D images an external labeling is created to name
the model's features. Hence, we obtain a temporal sequence of external
labelings; we call that sequence a \emph{dynamic labeling.} In
contrast, we call a single labeling in that sequence a \emph{static
  labeling}. Hence, we can distinguish models for \emph{static external
  labeling} that aim at individual labelings, and models for
\emph{dynamic external labeling} that aim at sequences of
labelings. Models on dynamic external labeling may define
constraints enforcing temporal coherence between consecutive labelings
in order to avoid distracting effects such as flickering and jumping
labels. In most of the literature on labeling of 3D models the labels are placed in the projection space, which is also our default here, but there are also some papers that place labels in the object space.

\subsection{Optimization problem}\label{sec:model:optimization}
Next we introduce the formal definition of the external
labeling problem.  As input we are given a drawing area $D=(R,A)$
partitioned into the image region $R$ and the labeling region $A$, a
feature set $F$ and a set $M$ of model parameters to be specified
below. We call the tuple $I=(D,F,M)$ an \emph{instance} of external
labeling. 
A labeling $\mathcal L$ of $F$ that satisfies all model
parameters of $M$ is called a \emph{feasible} labeling of $I$. We
denote the set of all feasible labelings of $I$ by $\mathcal S$.

The model parameters are a set of layout-specific criteria
that must be satisfied. Typical parameters are the selected leader
types, site, port, and reference point constraints, and hard constraints
that restrict the mutual interplay of the callouts or their
interference with the illustration in the background layer. While the
leader types have been specified in Section~\ref{sec:model:features}, the site, port, and
reference point constraints define the admissible placement of the site (fixed, sliding, free), of the port (fixed, sliding), and the
reference point (fixed, sliding, free). Examples of hard constraints are
crossing-free leaders, non-overlapping labels, and avoidance of
obstacles in the illustration. The actual set of model parameters
strongly depends on the application and is determined by the
designer, see Section~\ref{sec:visual-aspects} for some commonly used guidelines.

Most commonly, external labeling is seen as an optimization problem to
find the best among all feasible labelings (or at least a reasonably good one). Hence, it remains to
define a cost function $c\colon \mathcal S \to \mathbb R^+$ that
measures the quality of the feasible labelings in~$\mathcal S$, e.g., the total length of all leaders. For a detailed discussion on typical optimization criteria see Section~\ref{sec:visual:placement}.

\begin{problem}[\textsc{ExternalLabeling}]~\\~
  \label{problem:external-labeling}
\begin{tabular}{ll}
  \textbf{Input:} & Instance $I=(D,F,M)$, cost function $c\colon \mathcal S \to \mathbb R^+$\\
  \textbf{Output:} & Optimal labeling $\mathcal L\in \mathcal S$, i.e., $c(\mathcal L)\leq c(\mathcal L') \forall \mathcal L'\in \mathcal S$.                   
\end{tabular}
\end{problem}

\section{Visual Aspects of External Labeling}\label{sec:visual-aspects} 

\begin{figure}[t]
  \centering
  \begin{subfigure}[b]{0.49\linewidth}
    \centering
    \includegraphics[page=1]{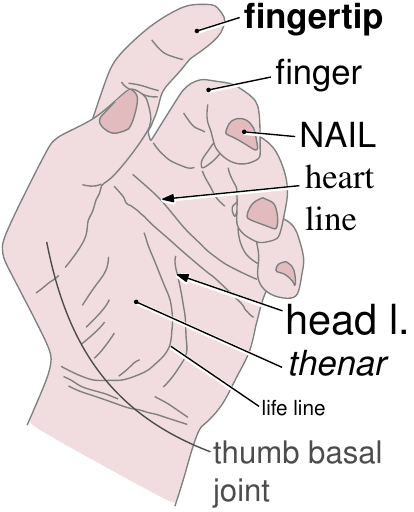}
    \subcaption{\label{fig:visual-aspects:general}}
  \end{subfigure}
  \hfill
   \begin{subfigure}[b]{0.49\linewidth}
    \centering
    \includegraphics[page=2]{fig/visual-aspects}
    \subcaption{\label{fig:visual-aspects:placement}}
  \end{subfigure}
  \caption{(\subref{fig:visual-aspects:general})~Different examples of rendering and typesetting callouts.
    (\subref{fig:visual-aspects:placement})~The labels on the
    right-hand side (blue) comply with Criteria C1--C5, the labels on
    the left-hand side (red) do not.}
\end{figure}

In this section we discuss various visual aspects that affect the aesthetic quality and usability of external labeling. 
In general, it is far from obvious, which type of external labeling is the best choice and this highly depends on the application. 
We discuss several important criteria that must be taken into account when deciding on a type of external labeling. We distinguish between the
\emph{general style} of the labeling and the actual
\emph{placement} of the labels based on it. 

\subsection{General Style}

The visual appearance of a callout in an externally labeled illustration depends on many aspects, which influence the legibility and readability, as well as the overall aesthetics of the illustration to various degrees.
While the focus of this survey is on the algorithmic aspect of optimizing the positioning of labels and leaders, this section gives a high-level overview of other practically important parameters affecting the quality of a labeling. Figure~\ref{fig:visual-aspects:general} illustrates the visual effects of these design choices. 

First of all, most of the leaders used in external labeling algorithms are either straight-line leaders (see Section~\ref{sec:star:sl}) or polylines whose segments have restricted slopes (see Section~\ref{sec:star:poly}). Only few papers also considered curved leaders~\cite{Kindermann2014}. 
The choice of the leader type has a strong influence on the overall appearance. In terms of readability, it is important to consider the overlay of the illustration in the background layer and the leader curves in the labeling layer. One must ensure that lines in the background layer and leader lines can be easily differentiated~\cite{r-tsi-17}, which means that in illustrations with arbitrary and irregular shapes (e.g., maps or anatomy drawings such as Fig.~\ref{fig:external-labeling} and~\ref{fig:visual-aspects:general}), straight and polyline leaders are preferable, whereas in illustrations with many straight lines of regular slopes (e.g., floorplans), curved leaders or straight and polyline leaders with non-aligned slopes should be considered~\cite{Wu2012}.

Richards~\cite{r-tsi-17} presented a list of best-practice guidelines originally for labeling black-and-white technical drawings in aerospace engineering, but most of them should also generalize to other domains. Leader lines should be straight wherever possible and they should consist of a black line and right next to it, on one side, a thin white line to separate the leader from the background. Moreover, leader lines (for area features) should terminate with a terminator symbol, either a small dot if the feature is large enough, or an arrowhead for small features. According to Richards~\cite{r-tsi-17}, polyline leaders producing staircase effects, possibly for aesthetic reasons, are acceptable as long as there are not too many leaders, but should be avoided for contexts where safety or efficiency of use are relevant.

The graphical rendering of leader lines themselves is also of relevance for the visual quality of a labeled image. Wu and Dalal~\cite{wd-plqm-05} considered different quality feature of rendering lines on electronic displays, including the raggedness, blur, waviness, and lumpiness of lines, depending on the method of digitizing Euclidean straight lines and using anti-aliasing effects. 

In addition to the leader, the label itself needs to be properly configured and displayed. Foremost, the typesetting of the label text must be done appropriately. For example, Degani~\cite{d-tfd-92} gave a detailed list of guidelines and principles for typesetting of flight-deck documentation, whose use of text is comparable to labeling. We mention the most important ones. While one can choose from a large variety of fonts, which should match the context of the illustration, for labels and short annotations generally sans-serif fonts are more legible than serif fonts. The selected font size depends on the estimated viewing distance and can be calculated according to standard guidelines. The same is true for non-textual labels such as icons. Lower-case words are more legible than upper-case words. Regular face is more legible than italic or bold face, which should only be used for highlighting specific words. Finally, a black font over white or lightly colored background has the best legibility. 
For the text itself, one consideration is whether abbreviations or line breaks should be used, e.g., Shimabukuro and Collins~\cite{sc-atld-17} presented a method for on-demand abbreviations of text labels.

\subsection{Placement}\label{sec:visual:placement}
From a visual point of view the general labeling style surely
influences the appearance of the image most. However, once the
labeling style has been fixed, the actual placement of the labels
comes to the fore.  Choosing the most legible and visually pleasing one among all feasible
placements is a complex task that requires highly elaborated
algorithmic techniques when done automatically, and a good balance of
possibly contradicting placement criteria.

\paragraph*{Static Labelings}
For static labelings we have identified the following decisive criteria based
on literature that justifies them by interviews with experts and manual analysis of
labeled drawings; see Tables~\ref{table:sl:properties} and~\ref{table:poly:properties}  for the according references.
\begin{enumerate}
\itemsep0em 
\item[C1] \emph{The leaders have small length.} 
\item[C2] \emph{The number of leader crossings and label overlaps is small.}      
\item[C3] \emph{The labels mimic the shape of the image.}
\item[C4] \emph{The labels are distributed evenly.}
\item[C5] \emph{The directions of the leader segments comply with a set of preferred  directions.}
\item[C6] \emph{The leaders have a small number of bends.}
\end{enumerate}
Criteria C1 and C2 reduce visual clutter and
ambiguities and ease association between label and feature; a crossing- and overlap-free solution is often preferred
or even required~\cite{Ali2005,Bekos2007}. The importance of both criteria is strengthened by
the fact that the majority of research takes into account both; see Tables~\ref{table:sl:properties} and~\ref{table:poly:properties}.

Criteria C3--C6 aim at
labelings that blend in with the image. For example, evenly
distributed labels lead to a homogeneous appearance of the labeling,
and labels mimicking the shape of the figure repeal the clear
distinction between labeling layer and background layer. They are justified in multiple references (e.g.,\cite{Ali2005,Mogalle2012,Niedermann2017}).
Concerning the number of occurrences in literature, the following
criteria are deemed less important, but still they are verified by
interviews with experts~\cite{Niedermann2017}.
\begin{enumerate}
\itemsep0em 
\item[C7] \emph{There is sufficient space between leaders.} 
\item[C8] \emph{Labels consist of single text lines if possible.}
\item[C9] \emph{Labels that are semantically related are grouped.}
\end{enumerate}
Criterion C7 reduces ambiguities between leaders, while Criteria C8
and C9 aim at a better legibility and comprehensibility of the
labeling, respectively. 

We stress that there is not a clear, empirically confirmed
ranking among these criteria, which is one of the research challenges
discussed in Section~\ref{sec:discussion}; their respective relevance and relative
priorities also depend on the application domain or even on the image
to be annotated. Still they are supported by general labeling and
drawing principles such as proximity, planarity, straightness/low
detour, and small ink-ratio
\cite{Imhof1975,Formann1991,p-wagehu-97,wpcm-cmga-02,Tufte2001}.
We also stress that some of these criteria might be in conflict with
each other. As a result, a good labeling is usually a compromise of
different criteria. To this end, one usually scales the different
criteria by suitable weight factors to obtain a multi-criteria cost
function $c$ in
Problem~\ref{problem:external-labeling}.

\paragraph*{Dynamic Labelings}
Dynamic labelings are especially found in view management systems, in
which the user can interactively explore 3D models. Hence, the
placement of the labels becomes an \emph{online} problem for which the
labeling algorithm needs to create a dynamic labeling without exactly
knowing the upcoming changes. As the frames of a dynamic labeling form
a sequence of static labelings, the criteria for the static case also
apply for the dynamic case, but are extended by the general
requirement of temporal coherence between two consecutive
frames. Following general dynamic map labeling
principles~\cite{Been2006}, \emph{flickering} and \emph{jumping}
labels should be prevented, while optimizing and preserving the
placement criteria listed above. The proposed approaches found in
literature can be categorized into two groups~\cite{Madsen2016}. The
approaches of the first group consider continuously moving labels,
which clearly prevents flickering and jumping labels. However, the
placement criteria such as overlap-free labelings and sufficient
spacing between labels, cannot be completely maintained. In contrast,
the approaches of the second group use \emph{hysteresis}, i.e., the
displacement of the labels is delayed to stabilize the motion of the
labels and to enforce selected placement criteria. However, this
easily leads to labels that jump from one position to the
next. Preserving the order of the labels and animations visualizing
the changes are possible tools to soften this effect. In general,
animations have been considered as an important tool in interactive
visualizations to visualize the transition from one state to the
other~\cite{Elmqvist2011}. We further note that Madsen et
al.~\cite{Madsen2016} showed in a detailed user study on labelings in
augmented reality that the participants performed better in the
conducted tasks when the labels were directly placed in the 3D space of
the object instead of the 2D space.  This result is further supported
by the user study by Peterson et al.~\cite{Peterson2009} showing that
incorporating depth information into the labeling increased the
performance of the participants in selection tasks. They particularly
showed that this technique softens the negative effects of overlapping
labels.  Still, only few approaches~\cite{Pick2010,Tatzgern2014} that
aim at a clear spatial distinction between objects and placed labels have been
developed for external labeling in object space so far.

\section{State of the art}\label{sec:star}

In this section we
describe a comprehensive collection of the state of the art on
external labeling algorithms. We discuss in total \arabic{allReferences}
references, which we collected in two main steps. Starting from a list
of references that we have collected over the years, we enriched this
list by a careful search in Google Scholar using different search
terms closely related to external labeling. Among others we used the
following terms: \emph{external labeling}, \emph{boundary labeling},
\emph{contour labeling}, \emph{excentric labeling}, \emph{callouts},
\emph{labeling with leaders}, \emph{external labels}, \emph{labeling
  with focus regions}, \emph{hyperleaders}, \emph{automatic text
  annotations}, \emph{label layout}, \emph{many-to-one labeling},
\emph{labeling of visualizations}.  We further enriched the list by
parsing the references and the citations of each of the
papers that we obtained in the first step as well as by taking into
account existing surveys. The result of our search consists of
references from journals and conference proceedings related to the
visual computing community (including information visualization,
geovisualization, augmented reality and human computer interaction)
and related to the algorithms community (including computational
geometry and graph drawing). We stress that our search was 
limited to scientific results and did not go beyond this area (e.g.,
to patents). The goal
of this section is to acquaint the reader with the state of the art
in both communities and to intensify the exchange of ideas
between theory and practice.

In order to structure this section, we group the results by their
distinctive features.  At the top level we use the leader type as
distinctive feature, because it substantially determines the visual
appearance of an external labeling. We have identified \slAll references
using straight-line leaders (Section~\ref{sec:star:sl}) and
\polyAll references using polyline
leaders (Section~\ref{sec:star:poly}), which partitions the results into two approximately equally sized
groups. At the same time this roughly groups the results into
practical and theoretical ones, which can be of more interest for
researchers in the visual computing and in the algorithms communities,
respectively. We further review the current state on curved
leaders comprising only few references
(Section~\ref{sec:star:curved-leaders}), and empirical evaluation
techniques in external labeling (Section~\ref{sec:eval-techniques}).

We note that our taxonomy as defined in Section~\ref{sec:model} does not
  include all the labeling models that use labels with leaders in general.
  It requires that the labels in the labeling region $A$ are clearly separated from the
  underlying illustration in the image region $R$. 
  Note that there exist labeling models in the literature using \emph{internal} labels with leaders (e.g.~\cite{Bell2001,Azuma2003,Maass2006,Schwartges2015}),
  where  this clear spatial separation
  between labels and illustration is not required.
Since such
  labeling models raise very different algorithmic problems due to the strong interference between the image and the labels, we exclude their
  discussion from this survey.

\subsection{Straight-line leaders}\label{sec:star:sl}

\begin{table*}[p]
  \centering
  \resizebox{\textwidth}{!}{

\begin{tabular}{|l|ababababababab|abababa|babababa|r|}
\hline& \multicolumn{14}{c|}{Contour Labeling}& \multicolumn{7}{c|}{Free Labels}& \multicolumn{8}{c|}{Non-Strict}&$\sum$\\
& \rcite{Bekos2004}& \rcite{Ali2005}& \rcite{Bruckner2005}& \rcite{Hartmann2005}& \rcite{Gotzelmann2006}& \rcite{Cmolik2010}& \rcite{bsfrm-rmsvmed-11}& \ocite{Fink2012}& \ocite{Haunert2014}& \rcite{Kindermann2014}& \rcite{Fink2016}& \rcite{Niedermann2017}& \rcite{bnn-rlbl-18}& \rcite{cb-relgv-18}& \ocite{Fekete1999}& \rcite{Gotzelmann2006B}& \rcite{Fuchs2006}& \rcite{Vollick2007}& \rcite{wtly-zapalmm-11}& \ocite{Heinsohn2014}& \rcite{Gemsa2015b}& \rcite{Hartmann2004}& \rcite{Stein08}& \rcite{Muhler2009}& \rcite{Pick2010}& \rcite{Mogalle2012}& \rcite{Tatzgern2013}& \ocite{Balata2014}& \rcite{Tatzgern2014}& 29\\
\hline \textbf{S1 labeling contour}&$\times$&$\times$&$\times$&$\times$&$\times$&$\times$&$\times$&$\times$&$\times$&$\times$&$\times$&$\times$&$\times$&$\times$&&&&&&&&&&$\times$&&$\times$&&&&16\\
S1.1 rectangle&$\times$&$\times$&&$\times$&&$\times$&&&&$\times$&$\times$&&$\times$&&&&&&&&&&&&&$\times$&&&&8\\
S1.2 circle&&$\times$&&$\times$&&$\times$&$\times$&$\times$&$\times$&&&&&&&&&&&&&&&&&&&&&6\\
S1.3 convex hull&&$\times$&$\times$&$\times$&&$\times$&&&&&&$\times$&&$\times$&&&&&&&&&&&&&&&&6\\
S1.4 other hull&&&&&$\times$&&&&&&&&&&&&&&&&&&&$\times$&&&&&&2\\
\hline\textbf{S2 site placement}&&$\times$&&$\times$&$\times$&$\times$&&&&&&&&$\times$&&$\times$&&$\times$&&&&$\times$&&$\times$&$\times$&&$\times$&&$\times$&12\\
\hline \textbf{S3 drawing criteria}&&&&&&&&&&&&&&&&&&&&&&&&&&&&&&\\
S3.1 crossing-free callouts&$\times$&$\times$&$\times$&$\times$&$\times$&$\times$&$\times$&$\times$&$\times$&$\times$&$\times$&$\times$&$\times$&$\times$&&&$\times$&&&&$\times$&&$\times$&$\times$&&&&&&18\\
S3.2 leader length&$\times$&$\times$&$\times$&$\times$&$\times$&$\times$&$\times$&$\times$&$\times$&&$\times$&$\times$&$\times$&$\times$&&$\times$&$\times$&$\times$&$\times$&$\times$&$\times$&$\times$&$\times$&$\times$&$\times$&$\times$&$\times$&$\times$&$\times$&27\\
S3.3 label spacing&$\times$&$\times$&&$\times$&$\times$&$\times$&$\times$&&&$\times$&&$\times$&$\times$&&$\times$&&&$\times$&&$\times$&$\times$&$\times$&&&&&$\times$&$\times$&&16\\
S3.4 leader direction&&$\times$&&$\times$&$\times$&$\times$&&$\times$&$\times$&&&$\times$&&$\times$&&&&$\times$&&$\times$&$\times$&$\times$&$\times$&&$\times$&&&&$\times$&15\\
S3.5 vert./horz. aligned labels&$\times$&&&&&$\times$&&&&$\times$&$\times$&&$\times$&&$\times$&$\times$&&$\times$&&$\times$&$\times$&&&&&$\times$&&&&11\\
S3.6 groups/clusters&&&&&$\times$&&&$\times$&&&&$\times$&&&&$\times$&&$\times$&&&&&&$\times$&&&$\times$&&&7\\
S3.7 leader-image occlusion&&&&&&&&&&&$\times$&&&&&&&&&&&$\times$&&$\times$&&&&&&3\\
S3.8 mixed labeling&&&&$\times$&$\times$&&&&&&&&&&&&&&$\times$&&&&&&&&&&&3\\
\hline\textbf{S4 temporal~coherence}&&$\times$&&$\times$&$\times$&$\times$&&&&&&&&&&$\times$&&&&&&&$\times$&$\times$&$\times$&$\times$&$\times$&$\times$&$\times$&12\\
\hline\textbf{S5 labeling techniques}&&&&&&&&&&&&&&&&&&&&&&&&&&&&&&\\
S5.1 init and improve&&$\times$&$\times$&$\times$&$\times$&$\times$&&&&&&&&$\times$&&&&$\times$&&$\times$&&$\times$&&$\times$&$\times$&&$\times$&$\times$&$\times$&14\\
S5.2 greedy&&&$\times$&&$\times$&$\times$&&&&&&&&$\times$&&&$\times$&&$\times$&$\times$&$\times$&$\times$&$\times$&&&$\times$&$\times$&&&12\\
S5.3 force-based&&$\times$&&$\times$&&&&&&&&&&&&&&&&$\times$&&$\times$&&&$\times$&&$\times$&$\times$&$\times$&8\\
S5.4 dynamic progamming&&&&&&&&$\times$&$\times$&&$\times$&$\times$&&&&&&&&&$\times$&&&&&&&&&5\\
S5.5 meta-heuristic&&$\times$&&&&&&&&&&&&&&&&$\times$&$\times$&&&&&&&&&&&3\\
S5.6 matching&$\times$&&&&&&&$\times$&&&&&&&&&&&&&&&&&&&&&&2\\
S5.7 plane sweep&&&&&&&&$\times$&&&&&&&&&&&&&&&&&&&&&&1\\
S5.8 mathematical prog.&&&&&&&&&&&&&$\times$&&&&&&&&$\times$&&&&&&&&&2\\
\hline\textbf{S6 contributions}&&&&&&&&&&&&&&&&&&&&&&&&&&&&&&\\
S6.1 implementation&&$\times$&$\times$&$\times$&$\times$&$\times$&$\times$&$\times$&$\times$&$\times$&&$\times$&$\times$&$\times$&$\times$&$\times$&$\times$&$\times$&$\times$&$\times$&$\times$&$\times$&$\times$&$\times$&$\times$&$\times$&$\times$&$\times$&$\times$&27\\
S6.2 analysis: exist. drawings&&$\times$&&$\times$&&&&&&&&$\times$&&&&&&$\times$&&&&$\times$&&&&$\times$&&&&6\\
S6.3 formal proofs&$\times$&&&&&&&$\times$&$\times$&&$\times$&$\times$&&&&&&&&&$\times$&&&&&&&&&6\\
S6.4 expert interviews&&&&&&&&&&&&$\times$&&$\times$&&&&&&&&&&$\times$&&$\times$&&&&4\\
S6.5 user study&&&&&&&&&&&&&$\times$&$\times$&$\times$&&&&&&&&&$\times$&$\times$&&&$\times$&&6\\
\hline\textbf{S7 algorithm type}&&&&&&&&&&&&&&&&&&&&&&&&&&&&&&\\
S7.1 heuristic&&$\times$&$\times$&$\times$&$\times$&$\times$&$\times$&&&$\times$&&$\times$&$\times$&$\times$&$\times$&$\times$&$\times$&$\times$&$\times$&$\times$&&$\times$&$\times$&$\times$&$\times$&$\times$&$\times$&$\times$&$\times$&24\\
S7.2 exact&$\times$&&&&&&&$\times$&$\times$&$\times$&$\times$&$\times$&$\times$&&&&&&&&$\times$&&&&&&&&&8\\
\hline\textbf{S8 community}&&&&&&&&&&&&&&&&&&&&&&&&&&&&&&\\
S8.1 algorithms&$\times$&&&&&&&&&$\times$&$\times$&&&&&&&&&&&&&&&&&&&3\\
S8.2 visual computing&&$\times$&$\times$&$\times$&$\times$&$\times$&$\times$&$\times$&$\times$&&&$\times$&$\times$&$\times$&$\times$&$\times$&$\times$&$\times$&$\times$&$\times$&$\times$&$\times$&$\times$&$\times$&$\times$&$\times$&$\times$&$\times$&$\times$&26\\
\hline\end{tabular}
  }
  \caption{Properties of the approaches for straight-line
    leaders. References are partitioned into three top-level groups by
    contour labeling, free label placement and non-strict external labelings. References considering excentric labeling are marked with ``$\excentricLabeling$''. }
  \label{table:sl:properties}
\end{table*}

Straight-line leaders are the simplest way to establish
the visual association between features and their external
labels. Further, they can be easily traced by the
reader~\cite{bnn-rlbl-18}. We first give an overview on the current research
(Section~\ref{section:star:sl:overview}) and then discuss the individual papers concerning their
particular contributions (Section~\ref{section:star:sl:details}).

\subsubsection{Overview}\label{section:star:sl:overview}
Using straight-line leaders for external labeling requires a broad
variety of design decisions. As a result plenty of different approaches
have been developed. We summarize the most important properties of
the approaches in Table~\ref{table:sl:properties} and discuss them in
the following.

\paragraph*{Label positions} The high-level grouping of the relevant
literature as seen in the top row of Table~\ref{table:sl:properties}
considers three general approaches of placing labels. The first group
of approaches considers contour labeling assuming that the labels are
placed along a pre-defined contour, which means that labels mimic the
silhouette of the illustration. Especially in professional
applications such as atlases of human anatomy domain experts require
this property~\cite{Niedermann2017}.  From an algorithmic point of
view, pre-defined contours have the advantage that they reduce the
search space by one dimension.  On the other hand, this also implies
that parts of the labeling region are entirely omitted for label
placement, while the chosen contour may not host all labels.  The
second group of approaches relaxes this problem by freely placing the
labels using the entire labeling area.  We distinguish between free
label placement within a continuous labeling
region~\cite{Fekete1999,Hartmann2004,Gotzelmann2006B,Vollick2007,Tatzgern2013,Balata2014,Heinsohn2014,Tatzgern2014},
and free label placement within a discretized labeling
region~\cite{Fuchs2006,wtly-zapalmm-11,Gemsa2015b}.  Finally, the last
group of approaches relaxes the requirement that labels may not be
placed inside the image region. As these approaches still place the
labels in the outer regions of the image as good as possible (in contrast to internal labelings with leaders), we
consider them to be in the corpus of literature on external labeling,
although they do not fully satisfy our definition of external
labeling; we call this placement technique \emph{non-strict} external
labeling accordingly.

\paragraph*{S1: labeling contour} For research assuming a pre-defined
labeling contour, the shape of the contour has an essential impact
both on the visual appearance of the labeling as well as on the
algorithmic properties. We distinguish between bounding
rectangles~(\slRectangleLabel), circles~(\slCircleLabel) and
(buffered) convex hulls~(\slConvexHullLabel) as possible shapes, but
in rare cases also more complex hulls~(\slOtherHullLabel) are taken
into account.

\paragraph*{S2: site placement} In illustrations
of human anatomy the locations of the sites are often pre-defined by a
domain expert before the labeling is
done~\cite{Niedermann2017}. However, for example, in interactive view
systems, this is often not adequate, and as a result the positions of the sites
need to be calculated. The sites should be placed on salient points of
the features such that they do not form dense clusters. Further,
considering a 3D object, the projection of the object into
the image space is not necessarily connected, but may consist of multiple
parts. Hartmann et al.~\cite{Hartmann2004} suggested to label the
biggest part or the part that promises the shortest leader.  In most
of the research that explicitly considers the placement of sites, this
is done in a pre-processing
step before the actual labels are placed~\cite{Hartmann2004,Ali2005,Hartmann2005,Gotzelmann2006B,Vollick2007,Muhler2009,Tatzgern2014}; only in two cases placing sites
and labels is considered
simultaneously~\cite{Cmolik2010,cb-relgv-18}. Hence, most of these
techniques can also be installed upstream for the other approaches
that require pre-defined sites. Moreover, some of the
approaches
also adapt the positions of the sites to improve temporal coherence in
dynamic scenarios~\cite{Ali2005,Hartmann2005,Gotzelmann2006,Muhler2009,Cmolik2010,Tatzgern2013}.

\paragraph*{S3: drawing criteria} 
All approaches pay special attention to intersections between
callouts, because they easily distract the reader and let the layout
appear cluttered; \slIntersection references guarantee that the 
created labeling is plane, while for the other
\transnum{\slIntersectionNotI} approaches intersections
(\slIntersectionLabel) might occur, e.g., if too many labels need to
be placed. Similarly, the length of the leaders (\slLengthLabel) is
considered to be an important drawing criterion to enforce unambiguity;
only \transnum{\slLengthNotI} references do not explicitly take leader
length into account. Moreover, the spacing between overlapping-free
labels (\slSpacingLabel) is considered by penalizing distances between
labels~\cite{Hartmann2004,Ali2005,Hartmann2005,Gotzelmann2006,
  Vollick2007,Tatzgern2013,Balata2014, Kindermann2014,Niedermann2017},
by stacking
them~\cite{Fekete1999,Gotzelmann2006B,Bekos2007,Ali2005,Hartmann2005,Heinsohn2014,Gemsa2015b},
or by distributing them evenly around the image
region~\cite{bsfrm-rmsvmed-11}.  Further, in~\transnum{\slDirectionI}references
the direction of the leaders~(\slDirectionLabel) is considered as
drawing criterion. Either the angles of the leaders are optimized with
respect to some pre-defined
directions~\cite{Hartmann2004,Gotzelmann2006,Gotzelmann2006B,Vollick2007,Stein08,Cmolik2010,Heinsohn2014,Gemsa2015b,Niedermann2017,cb-relgv-18},
the leaders radially emanate from a common
center~\cite{Ali2005,Hartmann2005,Fink2012,Heinsohn2014,Tatzgern2014},
or monotonically increasing angles are required when radially ordering
the leaders~\cite{Mogalle2012,Niedermann2017}. A vertical and
horizontal alignment of the labels or of groups of labels
(\slAlignmentLabel) is achieved by stacking the labels horizontally or
vertically
aligned~\cite{Fekete1999,Gemsa2015b,Gotzelmann2006B,Vollick2007,Heinsohn2014},
or by prescribing a rectangular labeling
contour~\cite{Bekos2007,Cmolik2010,Mogalle2012,Kindermann2014,Fink2016,bnn-rlbl-18}.
Further, the labels are either grouped or clustered (\slGroupingLabel), and occlusion of the image background by leaders
(\slObstaclesLabel) is taken into account.  Finally, \transnum{\slMixedI} references
consider mixed labelings (\slMixedLabel) supporting internal and
external labels.  Other criteria such as the distance between leaders
and sites are found in literature; for detailed lists of possible
criteria see for
example~\cite{Ali2005,Cmolik2010,Mogalle2012,Niedermann2017}.

\paragraph*{S4: temporal coherence}
In general all of the approaches can be used in dynamic scenarios, as
each displayed frame of an animation can be labeled
independently. However, this easily creates flickering and jumping
sites and labels. Therefore techniques that minimize the movement of
the sites and
labels~\cite{Ali2005,Hartmann2005,Gotzelmann2006,Stein08,Muhler2009,Cmolik2010,Pick2010,Tatzgern2013,Balata2014}
as well as hysteresis techniques that temporally freeze the current
labeling~\cite{Gotzelmann2006B,Muhler2009,Mogalle2012,Tatzgern2014,Balata2014}
have been developed to soften this effect.

\paragraph*{S5: labeling techniques}
The approaches for obtaining labelings with straight-line leaders are
rarely limited to one specific labeling technique, but they use
combinations of techniques to obtain an overall strategy.  In
particular, \slScheme references use the scheme of first computing an
initial labeling and then improving that labeling
successively~(\slSchemeLabel). We observe that greedy algorithms
(\slGreedyLabel) and force-based (\slForceBasedLabel) algorithms are
used by \transnum{\slGreedyI} and \transnum{\slForceBasedI}
references, respectively. They are easy to implement, run fast and
provide the possibility of integrating multiple drawing criteria.
Other labeling techniques such as dynamic programming (\slDPLabel),
meta-heuristic (\slMetaHeuristicLabel), matchings (\slMatchingLabel)
and sweep-line algorithms (\slSweepLabel) are rarely used. Fekete and
Plaisaint~\cite{Fekete1999}, Battersby et al.~\cite{bsfrm-rmsvmed-11}
Kindermann et al.~\cite{Kindermann2014}, and Götzelmann et
al.~\cite{Gotzelmann2006B} used highly specialized algorithms that do not match any of
the listed techniques; details are given in
Section~\ref{section:star:sl:details}.

\paragraph*{S6: contributions}
The contribution of research on the straight-line leaders is manifold.
Firstly, \slImp references describe implementations (\slImpLabel),
which are also used for evaluation purposes. All of them discuss
examples generated with the presented approaches.  Further, \transnum{\slMAI}
references justify their design decisions on the analysis of existing
drawings (\slMALabel), e.g., atlases of human anatomy and visual
dictionaries, and \transnum{\slEII} of them on interviews with domain experts
(\slEILabel) such as surgeons~\cite{Muhler2009},
radiologists~\cite{Mogalle2012} in the context of medical imaging,
designers and editors working on atlases of human
anatomy~\cite{Niedermann2017} as well as on interviews with experts of
related fields (user interface and graphics design)~\cite{cb-relgv-18}.
Further, \transnum{\slUSI} of them
additionally evaluate the resulting labelings by user studies
(\slUSLabel). Moreover, \transnum{\slProofsI} references give formal
guarantees on certain optimization criteria (\slProofsLabel), e.g.,
the total leader length. Apart from analyzing the quality of the
labelings, the performance of the presented approaches is discussed by
nine references in detail
\cite{Bruckner2005,Gotzelmann2006,Vollick2007,Cmolik2010,Fink2012,Gemsa2015b,Niedermann2017,cb-relgv-18}.

\paragraph*{S7: algorithm type}\ %
We consider two different types of
algorithms found in the literature. Most of the
references~(\slHeuristic) consider heuristics, i.e., approaches that
find good but not necessarily optimal solutions. Only
\transnum{\slExactI} approaches are exact algorithms in the sense that
they yield optimal solutions for pre-defined cost-functions.

\paragraph*{S8: community}\ %
We partition the research into references that either stem from the
algorithms community or the visual computing community. More precisely,
for each reference we take into account in which journal or conference
proceedings it has appeared and relate it to one of two communities
accordingly. We observe that for straight-line leaders the majority of
references (\slCommVis) stem from the visual
computing community, while only \transnum{\slCommAlgoI} references
come from the algorithms community.

\subsubsection{Detailed discussion}\label{section:star:sl:details}
In this part we discuss the research on straight-line leaders in
detail. To that end, we group the results with respect to the allowed
label positions, as apart from the chosen leader type the position of
the labels strongly determines the visual appearance of the labeling.

\paragraph*{Contour labeling}\ %
Bekos et al.~\cite{Bekos2004,Bekos2007} considered rectangles as
labeling contour and coined the term \emph{boundary labeling}. For
straight-line leaders they considered uniform labels on all four sides
of the rectangle and presented algorithms that find for $n$ sites a
plane labeling in $O(n\log n)$ time and a plane labeling minimizing
the total leader length in $O(n^{2+\delta})$ time for any $\delta> 0$;
for details refer to Sections~\ref{sec:techniques:sweep}
and~\ref{sec:techniques:weighted-matching}, respectively.  Kindermann
et al.~\cite{Kindermann2014} considered boundary labeling for
straight-line leaders in the context of annotating texts. They placed
the labels either in the left or right margin of the text. For uniform
labels they utilized the approach by Bekos et al.~\cite{Bekos2007}. In
contrast to $\lopo$-leaders running in between the text lines,
straight-line leaders intersect and easily clutter the annotated text.
Fink and Suri~\cite{Fink2016} presented dynamic programming approaches
that can deal with obstacles, which may not be crossed by
leaders. However, their approaches have asymptotically high running
times, e.g., for the one-sided case they obtained a running time of
$O(n^{11})$ and for the two-sided case they obtained a running time of
$O(n^{27})$ when minimizing the total leader length.  Barth et
al.~\cite{bnn-rlbl-18} used an integer linear programming formulation
to create one-sided boundary labelings with straight-line leaders (Section~\ref{sec:techniques:math-programming}). They used these
labelings in a user study on the readability of different
leader types. They particularly showed that straight-line leaders
perform well when the user is asked to associate labels with their
sites and vise versa. However, concerning the aesthetic preferences
the users favored $\lpo$-leaders and $\ldo$-leaders.

Ali et al.\cite{Ali2005} suggested circular contours and
contours that mimic the silhouette of the illustration (e.g., by a
buffered convex hull). In an initial procedure the labels are placed
along these contours such that they are stacked and no label--label
overlaps occur. Afterwards the positions of the labels are switched
until all leader--leader intersections are resolved. In a final
compaction step the requirement of being placed on a contour is
relaxed and the labels are pushed towards the illustration to reduce
the lengths of the leaders. In a companion paper Götzelmann et
al.~\cite{Gotzelmann2005} described a system for labeling 3D models
with internal and external labels. In this system the external label
placement is done by the approach by Ali et al.~\cite{Ali2005}.
Hartmann et al.~\cite{Hartmann2005} presented a high-level description of
that system. The work by Götzelmann et
al.~\cite{Gotzelmann2005} is not listed in Table~\ref{table:sl:properties} since 
it focuses on internal label placement, but the work by
Hartmann et al.~\cite{Hartmann2005} gives a high-level
description of the same system.

Battersby et al.~\cite{bsfrm-rmsvmed-11} proposed \emph{ring maps}
considering a circle as contours. Uniform labels are evenly
distributed around the circle such that each label is oriented with
the ray that emanates from the center of the circle and goes through
the center of the label. Hence, the labels are not axis-aligned but
radially aligned. The order of the labels around the circle is chosen
such that the resulting labeling is plane. Fink et al.~\cite{Fink2012}
also considered circles as contours in the context of excentric
labeling.  In contrast to related work~\cite{Fekete1999,Heinsohn2014},
the labels are required to touch the circle. For all presented
algorithms they gave formal guarantees on the drawing criteria. In
particular they presented dynamic programming approaches that support
general weighting functions. Haunert and
  Hermes~\cite{Haunert2014} considered a closely related setting using
  radial leaders but horizontal aligned labels. They reduced the
  problem to finding maximum weight independent sets in conflict
  graphs of the labels, which they solved by means of dynamic
  programming.

Similar to Ali et al.~\cite{Ali2005}, Čmolík and
Bittner~\cite{Cmolik2010} considered rectangles, circles, and convex
hulls as pre-defined labeling contours. For each area to be labeled
they generated a set of candidate callouts that differ in the
chosen sites and the chosen position of the labels on the
contour. Using a rating of the candidates based on fuzzy logic, they
utilized a greedy algorithm that selects a candidate for each
feature. In a successive work Čmolík and Bittner~\cite{cb-relgv-18}
adapted these techniques to label \emph{ghosted views} of 3D
models, i.e., views that use transparency to show occluded parts of
the models. They particularly focused on the placement of the sites on
salient parts of the model. They evaluated the approach in an
extensive user study. Similarly to the user study by Barth et
al.~\cite{bnn-rlbl-18}, the users were asked to associate labels with
their highlighted features. The stimuli were created by their approach
as well as by experts. 

Bruckner and Gröller~\cite{Bruckner2005} presented an approach for
labeling 3D models using a simple iterative algorithm that
places the labels along the convex hull of the projected
model. Overlaps are resolved by moving the labels along the contour
and by exchanging the positions of overlapping labels. If not all
overlaps can be resolved, a greedy algorithm selects the labels by
priority excluding overlapping labels. Intersections between callouts
are resolved in a post-processing step based on the approach presented
by Bekos et al.~\cite{Bekos2007}.
Niedermann et al.~\cite{Niedermann2017} used convex hulls to prescribe the
labeling contours of illustrations in atlases of human anatomy. They
presented a general dynamic programming approach that labels a given set
of sites with respect to a set of hard and soft constraints. They
proved that the calculated solution respects all hard
constraints and is optimal with respect to all soft constraints.  In
contrast to other dynamic programming approaches, the cost-function
not only supports the rating of single leaders, but also  of
consecutive leaders. Thus, they could incorporate
drawing criteria such as distances between labels and monotonically
increasing angles between leaders.

Götzelmann et al.~\cite{Gotzelmann2006} considered external labeling in
interactive 3D visualizations. They used the \emph{orbit} of a figure
as contour, i.e., a hull around the figure such that each point on the
hull has the same distance to the figure. They computed the initial
positions of the labels using a simple greedy algorithm. The movement
of the labels and the sites is controlled by a multi-agent system
such that for each label there are agents, i.e., local strategies,
that control the behavior of the labels. They paid
special attention to temporal coherency and combined their approach
with internal label placement.

\paragraph*{Free label placement}\ %
Fekete and Plaisant~\cite{Fekete1999} introduced \emph{excentric
  labeling}. For straight-line leaders they presented a simple approach
that stacks the labels on the left and right hand side of the focus
region either according to their vertical or to their horizontal
order. To improve temporal coherence, labels are only placed when the
focus region is not moved.
Heinsohn et al.~\cite{Heinsohn2014} also considered excentric labeling
for straight line leaders, but in contrast to Fekete and
Plaisant~\cite{Fekete1999}  the labels are displayed all the time.  Heinsohn et al.~suggested four approaches to place labels:
\begin{inparaenum}[(i)]
\item an approach that places all labels on a stack on the left hand side of the focus
region,
\item a radial approach, i.e., the leader of each label is
part of the ray that emanates from the center of the focus region and
goes through the point feature of the labels, 
\item a force-based
approach that prefers labels with radial leaders (see
Section~\ref{sec:techniques:force-directed}), and   
\item a
\emph{cake-cutting} approach that places the labels equally
distributed around the focus region.
\end{inparaenum}
  In case that not all overlaps
can be resolved, the focus region is shrunk to reduce the number of
point features within the focus region.

Götzelmann et al.~\cite{Gotzelmann2006B} extended the system by Hartmann
et al.~\cite{Hartmann2005} by 
supporting grouping of
labels. More precisely, for each group of features they first compute
the sites of the features and determine the centroid of those
sites. For each centroid they create rectangle candidates in the
labeling region that can host the grouped labels. Within a group the
labels are vertically aligned.  Vollick et al.~\cite{Vollick2007}
formalized the external labeling problem as an energy function that is
composed by non-linear terms each representing one drawing
criterion. The exact weighting of the terms was learned from existing
visualizations (e.g., from hand-made drawings) by applying
\emph{nonlinear inverse optimization}. The authors used simulated
annealing to find a locally optimal labeling from the energy function.

While those approaches make use of a continuous labeling region, the
following discretize the labeling region to find appropriate label
positions. Fuchs et al.~\cite{Fuchs2006} partitioned the labeling region
into small rectangular regions. Using a simple greedy algorithm the
labels were assigned to these regions such that the length of the
leaders is minimized. Apart from labeling all features of the
illustration, they also considered a \emph{labeling lens}, which is
conceptually identical to the focus region considered in excentric
labeling. In this case the same approach was applied but restricted to
that region. Further, the approach was also used for more general settings,
where the image region consisted of multiple parts.
Wu et al.\cite{wtly-zapalmm-11} used external labeling to annotate a
metro map with station names and photos. To that end, they underlied
the metro map with a fine raster describing the free and occupied
regions of the map. Only cells that are not occupied can be used for
the label placement. They applied a genetic algorithm  combined with a greedy
algorithm to determine the position of the labels.
In particular, they placed textual labels and image labels
independently allowing label--leader intersections between the two groups.
Gemsa et al.~\cite{Gemsa2015b} considered the special case of
panorama labeling, e.g., for labeling the skyline of a city. They showed that in the
proposed model it suffices to consider a finite set of candidates per
feature to cover the optimal intersection-free labeling. Further, they
presented a dynamic programming approach and mixed integer linear
programming formulations for computing optimal labelings with respect
to different cost functions and settings.

\paragraph*{Non-strict external labeling}\ %
Hartmann et al.~\cite{Hartmann2004} presented an approach for labeling
3D objects. As they considered complex geometric features their approach
first determines the location of the sites by shrinking the feature
until a single point remains. They argued that this point is an
appropriate choice of a non-convex object, because it is ``placed at a
visual dominant region''. For the actual labeling they presented an
approach that consists of two phases.  In the first phase 
an initial layout is computed using a force-based approach (see
Section~\ref{sec:techniques:force-directed}). In the second phase,
possible overlaps between labels are minimized using a combination of
the same force-based approach and a greedy algorithm. While an
external labeling is preferred, labels may overlap the image region.
Tatzgern et al.~\cite{Tatzgern2013} adapted the approach by Hartmann et
al.~\cite{Hartmann2004} to label 3D explosion diagrams. They first
computed an initial layout by forming clusters of labels with similar
texts, and placing each label using the first phase of the approach
presented by Hartmann et al.~\cite{Hartmann2004}. Then, in a second
step they selected for each cluster its best label with respect to a set
of optimization criteria. Afterwards, they applied the second phase of
Hartmann~et al.~\cite{Hartmann2004} to minimize overlaps. They applied
their algorithm in a dynamic 3D scene. However, they did not enforce a
strict separation between labeling and image region such that the labels could overlap the labeled 3D object. To ensure
temporal coherence, the order of the labels and sites is fixed during
the movement of the camera. Afterwards, when the camera stops possible
intersections are resolved.
Tatzgern et al.~\cite{Tatzgern2014} presented another force-based
approach for 3D objects, which they called \emph{hedgehog labeling}. The
labeling is done in 3D space using 3D poles instead of straight-line
leaders. The forces are used to resolve overlaps between labels. Again
the labels are not clearly separated from the labeled object, but the forces are defined such that labels occluding the image are avoided. Pick et al.~\cite{Pick2010}
  also placed the annotations within the 3D space. They placed the
  labels directly at their point features and used a force-based
  approach to adapt the labels dynamically. They did not explicitly
  exclude intersecting leaders, but they used forces that prevent label 
  occlusion. Further, they introduced a force that reduces oscillating
  movements of labels.

Stein and D{\'{e}}coret~\cite{Stein08} formalized the external labeling problem of
3D models in the 2D projection space as an energy function whose
minimum yields a labeling with short leaders, crossing-free
callouts, and labels that avoid the image region. They used a greedy
strategy to find a locally optimal labeling. They improved temporal
coherence by penalizing positions of labels that lie too far away from
the previous positions. They also avoided flickering
effects that occur when sites become invisible during interaction.

In the context of excentric labeling Balata et
al.~\cite{Balata2014} presented an approach for labeling moving unmanned
aerial vehicles. They did not explicitly distinguish between image and
labeling region, but they used a force-based approach that separates the labels
from their sites. This approach  ensures that leaders
do not intersect, but labels may overlap each other. To improve
temporal coherence they considered forces that soften distracting
flickering effects, and the possibility of temporally freezing 
labels. In a conducted user study they showed that freezing labels
outperformed the other approach in terms of users' error rates.

We observe that two approaches~\cite{Muhler2009,Mogalle2012} of
non-strict external labeling also considered labeling contours.  If
necessary (e.g., due to missing free space), Mühler and
Preim~\cite{Muhler2009} relaxed the requirement on placing labels along
a contour, but allowed the labels to overlap unimportant structures of
the image. To that end, they extended the approach of Ali et
al.~\cite{Ali2005} such that it also supports the annotation of ghost
views used in surgical planning.  Further, they conducted a user study
in which the participants were asked to judge 24 images using
different drawing styles. Among others, the user study showed that the
participants prefered grouped labels.  Mogalle et al.~\cite{Mogalle2012}
placed the labels inside of the image boundary such that they touch the
boundary.  They presented both a greedy algorithm that selects labels
candidates and an approach that shifts the labels along the
contours. In order to avoid overlaps with other pictorial elements
(e.g., legends) they relaxed the requirement that the labels are placed
along a rectangle. Instead, they integrated pictorial elements into the
shape of the contour.

\subsection{Polyline leaders}
\label{sec:star:poly}

Most of the polyline leaders in the literature have a schematic and tidy appearance caused by the restriction to segment orientations that are aligned with the Cartesian coordinate axes or possibly with their two bisecting diagonals. Moreover, almost all work on polyline leaders requires crossing-free leaders and considers the boundary labeling setting, where the contour $C_A$ is a rectangle.
Again, we first give an overview (Section~\ref{ssub:overview}) of the existing literature on labeling with polyline leaders in terms of several characteristic properties and grouped by different types of input specifications. This is followed by a more detailed discussion of the individual papers and their contributions (Section~\ref{sec:poly:details}).

\begin{table*}[p]
  \centering
  \resizebox{\textwidth}{!}{

\begin{tabular}{|l|ababababa|bab|aba|bab|abab|aba|r|}
\hline& \multicolumn{9}{c|}{Points}& \multicolumn{3}{c|}{Text}& \multicolumn{3}{c|}{Areas}& \multicolumn{3}{c|}{Mixed}& \multicolumn{4}{c|}{Dynamic}& \multicolumn{3}{c|}{X-to-1}&$\sum$\\
& \rcite{Bekos2007}& \rcite{Bekos2006b}& \rcite{Benkert2009}& \rcite{Bekos2008c}& \rcite{Bekos2010b}& \rcite{Huang2014}& \rcite{Kindermann2015}& \rcite{bckmm-blrd-18}& \rcite{bnn-rlbl-18}& \rcite{Lin2009}& \rcite{Kindermann2014}& \rcite{LinPTWY11}& \rcite{Bekos2010}& \rcite{Fink2016}& \rcite{YangDGM17}& \rcite{Loeffler2010}& \rcite{Bekos2011}& \rcite{Loeffler2015}& \ocite{Fekete1999}& \rcite{Ali2005}& \ocite{Bertini2009}& \rcite{Noellenburg2010}& \rcite{Lin2008}& \rcite{Lin2010}& \rcite{Bekos2015}& 25\\
\hline \textbf{P1 leader type}&&&&&&&&&&&&&&&&&&&&&&&&&&\\
P1.1 po-leaders&$\times$&$\times$&$\times$&&&$\times$&$\times$&$\times$&$\times$&&$\times$&&$\times$&$\times$&&$\times$&&&&$\times$&&$\times$&$\times$&&$\times$&15\\
P1.2 opo-leaders&$\times$&$\times$&&$\times$&&$\times$&&$\times$&$\times$&$\times$&$\times$&$\times$&$\times$&$\times$&&&$\times$&&&&&&$\times$&$\times$&&14\\
P1.3 do- and pd-leaders&&&$\times$&&$\times$&&&&$\times$&&&&&$\times$&$\times$&&&$\times$&&&&&&&&6\\
P1.4 other&&&&&&&&&&&$\times$&&&&&&&&$\times$&&$\times$&&&&&3\\
\hline \textbf{P2 objective}&&&&&&&&&&&&&&&&&&&&&&&&&&\\
P2.1 length minimization&$\times$&&$\times$&$\times$&$\times$&$\times$&$\times$&$\times$&$\times$&$\times$&$\times$&$\times$&$\times$&$\times$&&&&&$\times$&$\times$&&$\times$&$\times$&$\times$&$\times$&19\\
P2.2 bend minimization&$\times$&&&$\times$&&$\times$&&$\times$&&&&$\times$&&&&&&&&&&&&$\times$&&6\\
P2.3 multicriteria&&&&&&&&&&&&&&&$\times$&&&&$\times$&$\times$&$\times$&&&&&4\\
P2.4 other&&$\times$&$\times$&&&&$\times$&&&$\times$&&&&&&$\times$&$\times$&$\times$&&&&&$\times$&$\times$&$\times$&10\\
\hline \textbf{P3 contour}&&&&&&&&&&&&&&&&&&&&&&&&&&\\
P3.1 boundary 1-sided&$\times$&$\times$&$\times$&$\times$&$\times$&$\times$&&&$\times$&$\times$&$\times$&$\times$&&$\times$&$\times$&$\times$&$\times$&&&$\times$&&$\times$&$\times$&$\times$&$\times$&19\\
P3.2 boundary 2-sided (opposite)&$\times$&&$\times$&$\times$&$\times$&$\times$&&&&$\times$&$\times$&&$\times$&$\times$&&&$\times$&&&$\times$&&&$\times$&$\times$&$\times$&14\\
P3.3 boundary 2-sided (adjacent)&&&&&&&$\times$&$\times$&&&&&&&&&&&&&&&&&&2\\
P3.4 boundary 3-sided&&&&&&&$\times$&$\times$&&&&&&&&&&&&&&&&&&2\\
P3.5 boundary 4-sided&$\times$&&&&$\times$&&$\times$&$\times$&&&&&$\times$&&&&&&&&&&$\times$&&&6\\
P3.6 non-rectangular&&&&&&&&&&&&&&&&&&$\times$&$\times$&$\times$&$\times$&&&&&4\\
\hline \textbf{P4 label type}&&&&&&&&&&&&&&&&&&&&&&&&&&\\
P4.1 uniform labels&$\times$&$\times$&$\times$&$\times$&$\times$&$\times$&$\times$&$\times$&$\times$&&&$\times$&$\times$&$\times$&$\times$&$\times$&$\times$&$\times$&&$\times$&&$\times$&$\times$&$\times$&$\times$&21\\
P4.2 non-uniform labels&$\times$&$\times$&&$\times$&$\times$&$\times$&&&&$\times$&$\times$&$\times$&&$\times$&&&&&$\times$&&$\times$&$\times$&&&&12\\
\hline \textbf{P5 label positions}&&&&&&&&&&&&&&&&&&&&&&&&&&\\
P5.1 fixed positions&$\times$&&$\times$&&$\times$&&$\times$&$\times$&$\times$&&&$\times$&$\times$&$\times$&$\times$&&$\times$&&&&&&$\times$&$\times$&&13\\
P5.2 sliding positions&&$\times$&&$\times$&&$\times$&&&&$\times$&$\times$&&&$\times$&&$\times$&&$\times$&$\times$&$\times$&$\times$&$\times$&&&$\times$&13\\
\hline \textbf{P6 label ports}&&&&&&&&&&&&&&&&&&&&&&&&&&\\
P6.1 fixed ports&$\times$&$\times$&&$\times$&$\times$&$\times$&$\times$&$\times$&$\times$&$\times$&$\times$&$\times$&&$\times$&$\times$&$\times$&$\times$&$\times$&$\times$&$\times$&$\times$&$\times$&$\times$&$\times$&$\times$&23\\
P6.2 sliding ports&$\times$&$\times$&$\times$&$\times$&$\times$&$\times$&$\times$&$\times$&&$\times$&&$\times$&$\times$&&&$\times$&$\times$&&&&&&&&&13\\
\hline\textbf{P7 leader obstacles}&&&&&&&&$\times$&&$\times$&$\times$&&&$\times$&&$\times$&$\times$&$\times$&&&&&&&&7\\
\hline\textbf{P8 contributions}&&&&&&&&&&&&&&&&&&&&&&&&&&\\
P8.1 formal proofs&$\times$&$\times$&$\times$&$\times$&$\times$&$\times$&$\times$&$\times$&&$\times$&&$\times$&$\times$&$\times$&&$\times$&$\times$&$\times$&&&&$\times$&$\times$&$\times$&$\times$&19\\
P8.2 implementation&&$\times$&$\times$&&&&&&$\times$&$\times$&$\times$&&$\times$&&$\times$&&$\times$&&$\times$&$\times$&$\times$&$\times$&$\times$&$\times$&&14\\
P8.3 user study&&&&&&&&&$\times$&&&&&&$\times$&&&&&&$\times$&&&&&3\\
\hline \textbf{P9 algorithm type}&&&&&&&&&&&&&&&&&&&&&&&&&&\\
P9.1 exact&$\times$&$\times$&$\times$&$\times$&$\times$&$\times$&$\times$&$\times$&$\times$&$\times$&$\times$&$\times$&$\times$&$\times$&&$\times$&$\times$&$\times$&&&&$\times$&$\times$&$\times$&$\times$&21\\
P9.2 approximation&&&&&&&&&&&&&&&&&$\times$&&&&&&$\times$&&&2\\
P9.3 heuristic&&&&&&&&&&$\times$&&&&&$\times$&&&&$\times$&$\times$&$\times$&&$\times$&&&6\\
\hline\textbf{P10 techniques}&&&&&&&&&&&&&&&&&&&&&&&&&&\\
P10.1 dynamic programming&$\times$&$\times$&$\times$&&&$\times$&$\times$&$\times$&&$\times$&$\times$&$\times$&&$\times$&&$\times$&&$\times$&&&&&&&$\times$&13\\
P10.2 plane sweep&$\times$&&$\times$&&$\times$&&&&&&$\times$&$\times$&&&&&$\times$&&&&&$\times$&&$\times$&&8\\
P10.3 weighted matching&$\times$&&&&$\times$&&&&&&&&$\times$&&&&&&&&&&$\times$&&$\times$&5\\
P10.4 scheduling&&&&$\times$&$\times$&$\times$&&&&&&$\times$&&&&&&&&&&&&&&4\\
P10.5 init and improve&&&&&&&&&&&&&&&&&&&$\times$&$\times$&$\times$&&&&&3\\
P10.6 mathematical programming&&&&&&&&&$\times$&&&&&&$\times$&&$\times$&&&&&&&&&3\\
P10.7 meta heuristics&&&&&&&&&&$\times$&&&&&&&&&&&&&&$\times$&&2\\
P10.8 greedy&&&&&&&&&&&&&&&&&&&&&&&$\times$&&&1\\
P10.9 other algorithm&$\times$&&&&&&&$\times$&&&&&&&&&$\times$&&&&&&$\times$&&&4\\
P10.10 \NP-hardness&$\times$&$\times$&&$\times$&$\times$&$\times$&&&&&&$\times$&&$\times$&&&&&&&&&$\times$&&$\times$&9\\
\hline\textbf{P11 community}&&&&&&&&&&&&&&&&&&&&&&&&&&\\
P11.1 algorithms&$\times$&$\times$&$\times$&$\times$&$\times$&$\times$&$\times$&$\times$&&&$\times$&$\times$&&$\times$&&$\times$&$\times$&$\times$&&&&$\times$&$\times$&&$\times$&17\\
P11.2 visual computing&&&&&&&&&$\times$&$\times$&&&$\times$&&$\times$&&&&$\times$&$\times$&$\times$&&&$\times$&&8\\
\hline\end{tabular}
  }
  \caption{Properties of the approaches for polyline-line leaders. References are partitioned into six top-level groups by feature type (\emph{points}, \emph{text}, \emph{areas}), use of internal and external labels (\emph{mixed}), features changing over time (\emph{dynamic}), and many-to-one labeling (\emph{X-to-1}). References considering excentric labeling are marked with ``$\excentricLabeling$''.
  }
  \label{table:poly:properties}
\end{table*}

\subsubsection{Overview} 
\label{ssub:overview}

We first group the polyline-leader algorithms by their input specifications and then classify each paper along eleven properties as listed in Table~\ref{table:poly:properties} and briefly discussed in the following.

\paragraph*{Input specifications}
The high-level grouping of the relevant literature as seen in the top row of Table~\ref{table:poly:properties} first considers three types of static external, mostly boundary labeling problems. The majority of papers consider point feature labeling, some of which assume the points to be word positions in line-based text documents. While in the former case, labelings are usually unconstrained in the background and feature layers, text labeling often requires leaders that run horizontally  between two text lines and may use a track-routing area in a narrow strip between text and labels to move vertically. Another set of papers studies labelings of area features, in which each leader's site within the feature can still be chosen freely by the labeling algorithm. The next group of papers considers mixed labeling problems, which combine the placement of internal labels with external labels. Here internal labels form obstacles that restrict the feasible positions of the leaders. Next, the group of dynamic labeling algorithms take into account point features that move over time as the user may shift a focus region or change the zoom level. While a static labeling algorithm usually does not preserve temporal coherence when applied to dynamic inputs, an algorithm for a dynamic labeling problem, conversely, can easily be applied to a static input. Finally, three papers consider many-to-one (X-to-1) labeling problems, where a single label refers to multiple features.

\paragraph*{P1: leader type}
One of the primary properties to distinguish different labeling algorithms and their results is the used leader type. Most polyline leaders found in the literature use orthogonal segments and are either of type $\lp\lo$ (\plPOLabel) with a single bend inside the image region or they are of type $\lo\lp\lo$ (\plOPOLabel) with two bends in a narrow \emph{track routing area} outside the image region but inside the contour $C_A$. Several papers employ orthodiagonal leaders, usually of type $\ld\lo$ or $\lp\ld$ with a turning angle of $\pm 45^\circ$ (\plDOPDLabel). These are shorter than the respective $\lp\lo$-leaders and have been empirically confirmed to be aesthetically more pleasing, yet slightly less readable than $\lp\lo$-leaders~\cite{bnn-rlbl-18}. Few papers do not use fixed slopes for all polyline leader segments, but rather include segments of type $\ls$ between the site, the bends, and the label (\plOtherSlopeLabel). 

\paragraph*{P2: objective}
The most common objective function (see also Section~\ref{sec:visual:placement}) for polyline leaders is the total leader length (\plLengthLabel). A small leader length implies a short distance between site and label and thus optimizes the locality of labels. The criterion also minimizes the ink used to plot the leaders, and thus also the amount of overplotting of the image region. A number of papers further considers bend minimization as an objective (\plBendsLabel). Note that a leader without bends is usually also a shortest possible leader, whereas for leaders with bends no such claim about their length can be made. 
Further, some methods use a multicriteria objective function (\plMultiLabel) taking into account leader lengths, but also separation between callouts and displacement of features or labels~\cite{YangDGM17,Fekete1999,Ali2005,Bertini2009}.
Other objective functions (\plOtherLabel) are either arbitrary quality measures based on a single leader~\cite{Benkert2009}, finding a feasible labeling at all~\cite{Lin2009}, maximizing the number of labeled features~\cite{Kindermann2015}, maximizing the label size~\cite{Bekos2006b}, maximizing the number of internal labels in a mixed model~\cite{Loeffler2010,Bekos2011,Loeffler2015}, or minimizing the number of leaders in many-to-one labeling~\cite{Bekos2015,Lin2010,Lin2008}. 

\paragraph*{P3: contour}
Most of the literature on polyline leaders considers boundary labeling, where the contour is the bounding box of the image region $R$. The distinction within different algorithms is on the number of sides of $R$ that can be used for attaching labels. The simplest and most-studied case is 1-sided boundary labeling (\plOneLabel). From an algorithmic point of view the 1-sided problem reduces to finding the best ordering of the labels on that side. This is a less complex situation than multi-sided labelings, where one needs to determine a side assignment together with an ordering for the labels on each side. Interestingly, 2-sided settings having labels on two opposite sides (\plTwoLabel), say left and right of $R$, are well studied in \plTwo papers and admit (so far) more efficient algorithms than the case of labels on two adjacent sides (\plTwoAdjLabel) (or three sides (\plThreeLabel)) of $R$, say top and right, with only two references in the literature. 
In applications with labels being names or other short text strings, using the two opposite sides left and right of the image region is more natural, as the height of a horizontal text label is small and thus many such labels can fit. The top and bottom sides, however, offer space for much fewer labels as the widths of the labels quickly add up to the image width, unless the labels are rotated by $90^\circ$. Only if the labels have aspect ratio closer to 1, e.g., for icons or acronyms, or are rotated accordingly, then all four sides (\plFourLabel) are equally well suited for placing labels.
In fact, placing labels on all four sides of $R$ has been studied in six papers. Few papers study polyline leaders in the context of other contours (\plotherContourLabel) like circles in excentric labeling~\cite{Fekete1999,Bertini2009}, convex polygons~\cite{Loeffler2015,Ali2005}, or arbitrary silhouettes~\cite{Ali2005}.

\paragraph*{P4: label type}
We distinguish two types of labels. Uniform labels (\plUniformLabel) are the most commonly studied (\plUniform references)  and all have the same size (or at least the same height as found in single-line text labels) and thus can be represented as a set of empty boxes distributed along the contour. Algorithmically speaking one has to find a matching that assigns each feature to one of these empty boxes and the label text is guaranteed to fit into the box. The more general case of non-uniform labels (\plNonuniformLabel, \plNonuniform references) is usually more difficult to solve as the placement of the labels now depends on the ordering of the leaders around the contour, e.g., imagine a large label consuming more space and pushing neighboring labels aside. Many results on non-uniform labels comprise \NP-hardness proofs relating label orderings to job scheduling problems (see Section~\ref{sec:techniques:complexity}). Still, for a number of problem settings, efficient algorithms exist even for non-uniform labels~\cite{Lin2009,Kindermann2014,LinPTWY11,Noellenburg2010,Huang2014}. 

\paragraph*{P5: label positions}
The label positions are either fixed on the contour (\plFixedposLabel) or they can slide along it (\plSlidingposLabel). The former case is more restrictive but also simplifies the algorithmic problem in many cases as one degree of freedom is removed.
Fixed positions are mostly used for uniform labels (\plUniformLabel), since for non-uniform, individually sized labels fixed positions already dictate the assignment of features to labels. In combination with property \plFixedportLabel, fixed-position labels give rise to leaders whose geometry is fully specified once a feature is assigned to a label. On the other hand, in some situations sliding labels are needed to compute feasible labelings at all, e.g., when labels of non-uniform size can be stacked or when leaders are blocked from reaching certain label positions due to obstacles. 

\paragraph*{P6: label ports}
Label ports can either be at fixed positions (\plFixedportLabel) or they can slide along one side of the label box (\plSlidingportLabel). Again, fixed port positions (in combination with fixed reference points (\plFixedposLabel)) are algorithmically often easier to handle because the geometry of leaders is fully determined by the position of the site and the position of the fixed port. Moreover fixed ports, e.g., centered at the label, often produce tidier labelings as in consequence the distance between neighboring leaders is uniform. Sliding ports may produce shorter leaders, e.g., when connecting a site that sits below a label to the bottom corner of its label rather than to its midpoint. Many algorithms even work for both types of ports. 

\paragraph*{P7: leader obstacles}
Obstacles are areas in the image region that cannot be crossed by a leader. These can be important parts of the image, the individual words in a text document, or internal labels in case of mixed labeling. Obstacles are usually modeled as polygons. They restrict the options of connecting a feature to a label position and thus increase the difficulty of the algorithmic problem, e.g., they can make an instance infeasible.
A special situation arises in mixed labeling~\cite{Loeffler2010,Loeffler2015,Bekos2011}, where the algorithm can decide whether an internal label, i.e., an obstacle, is placed.

\paragraph*{P8: contributions}
We distinguish three (non-exclusive) types of results that are found in the literature on polyline leaders. Most of the collected work provides exact algorithms (see also property \plExactLabel) accompanied with formal proofs of correctness, optimality of the solution, and asymptotic running times (\plProofsLabel). Not all of these algorithms have been implemented and some even have running times that are too high to be of practical relevance. On the positive side, they provide certificates that the corresponding problems are polynomial-time tractable. Eight papers provide formally analyzed algorithms together with proof-of-concept implementations (\plImpLabel).
Some other papers do not contain formal results and instead report about implemented systems and algorithms for polyline leaders that are justified by case studies or experiments. Finally, three papers include results of user studies on the computed labelings (\plUserStudyLabel).

\paragraph*{P9: algorithm type}
Here we consider three different types of algorithms found in the literature. Most algorithms (\plExact references) are exact in the sense that they provide mathematically optimal solutions to their respective labeling problems (\plExactLabel). Approximation algorithms (\plApproxLabel), found in two papers, cannot guarantee to find optimal solutions, but they are accompanied by formal proofs that provide an approximation guarantee that quantifies how far a solution can be from the optimum in the worst case. Finally, heuristics (\plHeuristicLabel) may compute good and sometimes optimal solutions, yet no formal proofs and guarantees are provided in the respective papers.

\paragraph*{P10: techniques}
This property refers to the algorithmic techniques and principles used in the different papers. The main features and some generic descriptions of these techniques are sketched in more detail in Section~\ref{sec:techniques}. Most algorithms use dynamic programming (\plDPLabel), which is based on subdividing a problem instance into two (or more) independent subinstances and then creating a solution from compatible subsolutions. Especially 1-sided boundary labeling problems can also be solved by sweep-line algorithms (\plSweepLabel) that visit the sites and labels in a suitable sequence, say top to bottom, and compute the leaders at discrete events, e.g., when new points and labels are reached. The weighted matching technique (\plMatchingLabel) is well suited for problems optimizing over a finite set of possible site--label assignments modeled as a weighted bipartite graph, e.g., for instances with fixed label positions, fixed ports, and uniform labels.
A scheduling algorithm (\plSchedLabel) is used for the 1-sided labeling of collinear points~\cite{Bekos2008c} and the scheduling technique is also the basis for a number of \NP-hardness reductions~\cite{Bekos2010b,LinPTWY11,Huang2014}.
Further, some papers use global optimization techniques such as SAT solving~\cite{Bekos2011} and mathematical programming (\plILPLabel), which includes integer linear programming~\cite{bnn-rlbl-18,Bekos2011} and quadratic programming~\cite{YangDGM17}. Other papers use meta heuristics (\plMetaLabel) such as genetic algorithms~\cite{Lin2009} and simulated annealing~\cite{Lin2010} or heuristics that first construct an initial solution and then improve it by local modifications (\plLocalLabel)~\cite{Fekete1999,Bertini2009,Ali2005}.
A greedy algorithm (\plGreedyLabel) is used for a crossing-minimization heuristic in many-to-one labeling~\cite{Lin2008}.
Among other techniques (\plOtherTechLabel) are algorithms that resolve leader crossings by swapping some site--label assignments that do not affect the sum of leader lengths~\cite{Bekos2007}, recursive approaches~\cite{Bekos2011}, and a reduction to finding weighted independent sets in outerstring graphs~\cite{bckmm-blrd-18}. 
Finally, some papers contain \NP-hardness reductions (\plComplexityLabel, see also Section~\ref{sec:techniques:complexity}), mostly for labeling problems with non-uniform labels. These are used to justify heuristics and approximation algorithms or simplifications of the model, e.g., the use of uniform labels.

\paragraph*{P11: community}
This property is identical to S8 for straight-line leaders. Interestingly, polyline leaders are studied primarily in the algorithms community (\plAlgLabel, \plAlg references), but also in several papers in the visual computing community (\plVisLabel, \plVis references).

\subsubsection{Detailed discussion} \label{sec:poly:details}

We start our discussion with two types of orthogonal leaders, namely $\lp\lo$-leaders (Section~\ref{sec:star:po}) and $\lo\lp\lo$-leaders (Section~\ref{sec:star:opo}). Then we discuss 1-bend orthodiagonal leaders  (Section~\ref{sec:star:do}).
Since the vast majority of the literature considers rectangular contours, i.e., the boundary labeling problem, we assume that the contour $C_A$ is an axis-aligned rectangle, unless stated otherwise.

\paragraph{$\lp\lo$-leaders}
\label{sec:star:po}
Orthogonal $\lp\lo$-leaders are the simplest and most commonly used polyline leader type in external labeling.
They are frequently found in the algorithmic literature, but also appear in professional information graphics and are implemented in some practical labeling methods.

\paragraph*{1-sided boundary labeling}
In the 1-sided setting, one of the simplest boundary labeling problems is to assign external labels of uniform size with fixed ports and thus fixed reference points aligned on a single, say vertical, side of the image region $R$ to a set of $n$ point features in $R$ such that the $\lp\lo$-leaders are crossing-free. The most commonly used objective function is to minimize the total leader length. For this problem Bekos et al.~\cite{Bekos2007} presented an $O(n^2)$-time algorithm to find a length-minimal solution. Their algorithm runs in two phases. First, sites and labels are matched so that they have the same vertical order. Second, leader crossings are iteratively resolved without changing the total leader length. Later Benkert et al.~\cite{Benkert2009} presented an algorithm based on the sweep-line paradigm that visits all sites and labels in their vertical order. Whenever a new label becomes available, a site is selected such that their leader is guaranteed to be crossing-free and the total length remains minimum. Their algorithm improved the running time to $O(n \log n)$. A very different method to compute length-minimal solutions is a simple integer linear program (ILP) proposed by Barth et al.~\cite{bnn-rlbl-18}. This approach generalizes to arbitrary leader shapes and label positions, as it basically computes a minimum-length crossing-free perfect matching between the given sites and fixed label positions. While in general ILP solving does not scale well, it is still a practical approach for smaller instances. Nöllenburg et al.~\cite{Noellenburg2010} then relaxed the restriction to fixed reference points with uniform labels and allowed sliding reference points and non-uniform labels. They presented another sweep-line algorithm that computes a length-minimal and crossing-free solution for this problems in $O(n \log n)$ time. Huang et al.~\cite{Huang2014} additionally relaxed the constraint to have fixed-port labels and allowed sliding ports. In this model, they proposed an $O(n^3)$-time algorithm for minimizing the total leader length.
A heuristic framework for external labeling has been proposed and implemented by Ali et al.~\cite{Ali2005}. They considered rectangular contours as well as other silhouettes and use rectilinear leaders as one option. The algorithm first determines suitable label positions, e.g., by projecting each site to the closest point of the contour or using force-based methods, and then tries to iteratively resolve leader crossings until a crossing-free solution is found. While this can produce reasonable labelings, no guarantees on optimality or termination are given.

Fink and Suri~\cite{Fink2016} proposed a collection of dynamic programming algorithms for boundary labeling with obstacles. Each obstacle is modeled as a rectilinear polygon and no leader may cross any obstacle. With their algorithm for 1-sided boundary labeling with $\lp\lo$-leaders and uniform labels, they could minimize the total leader length in $O(n^4)$ time if label positions are fixed and in $O(n^7)$ time otherwise. Their method also extends from point to area features at the cost of another linear factor. If, however, the labels have non-uniform size, they proved \NP-hardness for deciding the existence of a crossing-free solution. A more restricted case of boundary labeling with obstacles has been studied by Löffler and Nöllenburg~\cite{Loeffler2010}. They made the simplifying assumptions that all obstacles are congruent rectangles (which may model a set of internal labels). They provided dynamic programming algorithms for placing non-crossing $\lp\lo$-leaders such that the number of intersected obstacles is minimized. For some cases, e.g., when the obstacles can intersect each other or leaders may cross, they also proved \NP-hardness results.

For objective functions other than length minimization Benkert et al.~\cite{Benkert2009} proposed a general $O(n^3)$-time algorithm using a dynamic programming approach 
that optimizes the labeling using uniform, fixed-position labels with sliding ports. Kindermann et al.~\cite{Kindermann2014} generalized this algorithm to instances with non-uniform labels using rasterized label positions, where one label may occupy multiple raster slots. This increases the running time to $O(n^4 m^3)$ and the memory space to $O(n^3 m^2)$, with $m>n$ being the number of raster slots. Kindermann et al.\ used this algorithm for text labeling and made some adaptations to preferably route the $\lp$-segments between two lines of text rather than striking them out.
Another interesting variation is 1-sided multi-stack labeling~\cite{Bekos2006b}, where labels take positions in two or more label stacks on the same side of $C_A$. However, using $\lp\lo$-leaders the authors only showed that for non-uniform labels it is \NP-hard to decide the existence of a crossing-free two-stack labeling. Feasible algorithms in this model, e.g., for uniform labels, are missing.

Finally, two papers considered many-to-one boundary labeling, where sets of point features can share the same label. Lin et al.~\cite{Lin2008} considered individual $\lp\lo$-leaders and labels with multiple ports, one for each feature with that label. They proved that minimizing crossings in 1-sided many-to-one boundary labeling with $\lp\lo$-leaders is \NP-hard, but they also presented a greedy heuristic iteratively assigning label positions with locally fewest leader crossings. Bekos et al.~\cite{Bekos2015} investigated a different labeling style using hyperleaders (recall Section~\ref{sec:model:features}).
Here a $\lp\lo$-hyperleader for $k$ sites consists of $k$ vertical $\lp$-segments connecting the sites to a single horizontal $\lo$-segment, which connects to the label. 
They considered three problem variants. For crossing-free labelings, the problems are to minimize the number of labels or the total leader length, and for labelings with crossings, the problem is to minimize their number. Both problems with crossing-free leaders can be solved in polynomial time using dynamic programming. Crossing minimization can be solved in $O(n\cdot c)$ time, where $n$ is the number of sites and $c$ is the number of labels, if the label order is given; if the order is unconstrained, the problem becomes \NP-hard.

\paragraph*{2-sided boundary labeling}
A natural extension of the 1-sided case is to admit 2-sided placement of labels, either on two opposite or two adjacent sides of the bounding rectangle. This makes the algorithmic optimization problem more complex, as the distribution of labels to either side becomes an additional degree of freedom.

For uniform labels with fixed positions on two opposite sides of $C_A$, Bekos et al.~\cite{Bekos2007} presented an $O(n^2)$-time algorithm for computing a length-minimal solution. Using dynamic programming (Section~\ref{sec:techniques:dynamic-programming}) they obtained a length-minimal assignment of sites and labels, from which crossings were resolved independently on both sides as no pair of leaders to opposite sides can intersect in a length-minimal solution.
When replacing point features by a certain type of polygonal area features, Bekos et al.~\cite{Bekos2010} gave an $O(n^2 \log^3 n)$-time algorithm for length-minimal two-sided boundary labeling, in which the sites of the leaders are located along the sides of the area features. The difference of this algorithm is that the first step to compute the length-minimal site-label assignment uses the weighted matching technique (Section~\ref{sec:techniques:weighted-matching}). The crossing removal is done as before. For uniform labels with sliding positions, Huang et al.~\cite{Huang2014} presented an $O(n^5)$-time algorithm as an extension of their algorithm for the 1-sided case. The heuristic framework of Ali et al.~\cite{Ali2005}, already mentioned for 1-sided labeling, can be applied equally well to 2-sided instances.

For instances with obstacles, the dynamic programming algorithms presented by Fink and Suri~\cite{Fink2016} for minimizing the total length of a labeling with 1-sided $\lp\lo$-leaders can be extended to 2-sided instances, at the cost of a significant increase in time complexity from $O(n^4)$ to $O(n^9)$ for fixed label positions and from $O(n^7)$ to $O(n^{15})$ for sliding label positions.

For arbitrary objective functions, Benkert et al.~\cite{Benkert2009} generalized their dynamic-programming algorithm for 1-sided $\lp\lo$-leader labeling to two-sided boundary labeling, yet with a worst-case running time of $O(n^8)$. 
The reason for the algorithm's much higher complexity compared to $O(n^2)$ for length minimization~\cite{Bekos2007} is that for non-length-minimal solutions the sites can no longer be easily split into two independent subproblems for the two opposite sides.
The extension of Kindermann et al.~\cite{Kindermann2014} for non-uniform labels in 1-sided boundary labeling is not properly generalized to optimize 2-sided instances. Instead, for their text labeling application, the authors proposed to simply split the sites at the horizontal median position into two balanced parts and then run the 1-sided algorithm for each side independently.

Many-to-one labeling has also been studied for the 2-sided setting with labels on opposite sides. Since the crossing minimization problem considered by Lin et al.~\cite{Lin2008} remains \NP-hard for the 2-sided case, they again described a heuristic algorithm to obtain a labeling, in which all features sharing the same label connect with their own $\lp\lo$-leader to a single label box on one of the two opposite boundaries of $C_A$. Their algorithm constructs a weighted graph, in which they (heuristically) determine a vertex bisection of minimum weight, which induces the partition of labels to the two boundary sides. The weights are chosen such that the computed bisection has few leaders possibly crossing each other. Then each side is solved with their 1-sided algorithm. The hyperleader labeling by Bekos et al.~\cite{Bekos2015} is also considered in a 2-sided setting, but using duplicated labels that are vertically aligned on both sides. This actually makes the problem algorithmically simpler because now each hyperleader spans the entire width of the image region and splits the instance into two parts. Therefore the algorithms for crossing-free labelings with minimum number of labels or minimum length are improved to $O(n)$ and $O(n^2)$ running times, respectively. The crossing minimization problem that was shown to be generally \NP-hard for the 1-sided case remains open for 2-sided labelings. For a fixed label order, the polynomial-time 1-sided algorithm for crossing minimization can also be applied to the 2-sided case.

A different 2-sided setting is to have labels on two adjacent sides of $C_A$, e.g., the top and the right side. Here it is no longer the case that a crossing-free solution always exists. Kindermann et al.~\cite{Kindermann2015} presented an $O(n^2)$-time dynamic programming algorithm to compute a valid crossing-free solution if one exists, both for labels with fixed or sliding ports. Further, they presented an algorithm to maximize the number of labeled sites in a crossing-free labeling in $O(n^3 \log n)$ time. Finally, they modified their algorithm so that it can minimize the total leader length rather than computing some crossing-free solution, however, at the cost of an increase of the time complexity to $O(n^8 \log n)$ and a space bound of $O(n^6)$. Recently, Bose et al.~\cite{bckmm-blrd-18} improved this to an $O(n^3 \log n)$-time algorithm, again using the dynamic programming technique. An interesting algorithmic open problem is to investigate the remaining difference in the $O(n^2)$ running time for length minimization with labels on two opposite sides of $C_A$ and the $O(n^3 \log n)$ running time for labels on two adjacent sides.

\paragraph*{Multi-sided boundary labeling}

Three-sided labeling is explicitly considered only in two papers~\cite{Kindermann2015,bckmm-blrd-18}. Kindermann et al.~\cite{Kindermann2015} showed that this problem can be reduced to splitting an instance into two 2-sided subproblems. This results in an $O(n^4)$-time algorithm working in linear space for finding a crossing-free solution (if one exists) or for maximizing the number of labeled features. For 4-sided instances, they proposed an algorithm that considers all possibilities to split into two special 3-sided instances to be solved as before. This results in an increase of the running time to $O(n^9)$. Bose et al.~\cite{bckmm-blrd-18} improved upon this by presenting new algorithms for 3- and 4-sided boundary labeling that can even find length-minimal or bend-minimal solutions, for fixed and sliding ports and optionally with the leaders avoiding a set of obstacles. This is achieved by reducing the boundary labeling problem to a weighted independent set problem in outerstring graphs, for which a recently published algorithm is used~\cite{Keil2017}. 
From a heuristic perspective, again the labeling framework of Ali et al.~\cite{Ali2005} is flexible in terms of the contour $C_A$ and can thus be applied to place labels at three or four sides of a rectangle, as well as around a more complex silhouette contour of the image. 

\paragraph{$\lopo$-leaders}
\label{sec:star:opo}

Orthogonal $\lopo$-leaders have a slightly more complex shape than the corresponding $\lpo$-leaders since they contain an additional segment. As a result, the readability of the labelings obtained by using such leaders decreases with respect to the ones obtained by using $\ls$-, $\lpo$- or $\ldo$-leaders, as observed by Barth et al.~\cite{bnn-rlbl-18}. On the positive side, however, the central $\lp$-segment of each $\lopo$-leader lies outside the image region $R$ (which, for this particular type of leaders, is usually assumed to be rectangular, as otherwise the readability of the obtained labelings further decreases) routed within a so-called \emph{track routing area} that is wide enough to accommodate all $\lp$-segments. This implies that $\lopo$-leaders are particularly useful for text annotation purposes~\cite{Lin2009,Kindermann2014}, since the first $o$-segment of each $\lopo$-leader, which is the only one that interferes with the underlying text, can be drawn between the lines of the text.

\paragraph*{1-sided boundary labeling} In the $1$-sided setting, computing a feasible labeling, in which no two $\lopo$-leaders intersect, is an easy task. Assuming without loss of generality that the external labels are attached on a vertical side of the image region $R$, we observe that the vertical order of the sites must be identical to the vertical order of their corresponding labels. This observation directly gives rise to an $O(n \log{n})$-time constructive algorithm~\cite{Bekos2007}. For the problem of finding labelings, in which the total number of bends is minimized, Bekos et al.~\cite{Bekos2007} proposed an $O(n^2)$-time dynamic programming based algorithm to find bend-minimal solutions, assuming that the ports of the labels are sliding. For the case that the ports of the labels are fixed, Huang et al.~\cite{Huang2014} proposed a slightly less efficient algorithm (which is also based on dynamic programming) to find a bend-minimal solution in $O(n^3)$ time. In the context of collinear sites, Bekos et al.~\cite{Bekos2008c} proposed an $O(n \log{n})$-time algorithm to compute length-minimal solutions by employing a linear-time reduction to a single machine scheduling problem. Notably, their algorithm is not difficult to be adjusted to the more general $1$-sided boundary labeling setting assuming that the ports of the labels are sliding; for details refer to Section~\ref{sec:techniques:scheduling}. 

For the case that rectangular-shaped obstacles (e.g., internal labels) are allowed within the image region $R$, which must not be crossed by the leaders, Fink and Suri~\cite{Fink2016} presented an $O(n^{11})$-time algorithm to compute a length-minimal solution, under the assumption that the labels are of uniform size. Bekos et al.~\cite{Bekos2011} and L\"offler et al.~\cite{Loeffler2015} studied more general mixed labelings, in which each site can be labeled either with an internal label that is located directly to its top-right or with an external label through an $\lopo$-leader. When  labels are of uniform height and the external labels are to the right side of the image region $R$, Bekos et al.~\cite{Bekos2011} provided an $O(n \log{n})$-time algorithm for maximizing the total number of internal labels. Note that the case where  the external labels are to the left side of $R$ is not symmetric and it turns out to be more difficult. For this case, Bekos et al.~\cite{Bekos2011} provided a quasi-polynomial algorithm of $O(n^{\log{n}+3})$ time complexity, a more efficient $2$-approximation and an integer linear programming formulation for maximizing the number of internal labels. L\"offler et al.~\cite{Loeffler2015} presented an improved algorithm to solve the case, in which the external labels are on the left side of $R$ and they are connected to their corresponding sites with type-$\lo$ leaders, in time $O(n^3(\log{n}+\delta))$, where $\delta$ denotes the minimum of $n$ and the inverse of the distance of the closest pair of points.

$1$-sided boundary labelings with $\lopo$-leaders have also been studied in the many-to-one setting, in which a point can be attached to more than one label~\cite{Lin2008,Lin2010}. Lin et al.~\cite{Lin2008} observed that in this setting crossings will inevitably occur, and proved that minimizing their number is \NP-hard. Finally, they presented a $3$-approximation algorithm, which places the labels in ``median order'' as introduced by Eades and Wormald~\cite{EadesW94}. In a subsequent work, Lin~\cite{Lin2010} allowed more than one occurrence of the same label, in order to avoid crossings that negatively affect the readability of the produced labelings. Naturally, the focus of his work was on minimizing the total number of used labels; a task which in the 1-sided case can be solved in $O(n \log{n})$ time by a simple top to bottom traversal of the sites, assuming that the labels are along a vertical side of the image region~$R$.

We conclude this subsection by mentioning that $1$-sided boundary labeling with $\lopo$-leaders has also been studied when more than one stack of labels are allowed along the side $s$ of the image region $R$ containing the labels, under the additional assumption that the labels are all of uniform size and their ports are sliding~\cite{Bekos2006b}. In this setting, Bekos et al.~\cite{Bekos2006b} studied the problem of finding boundary labelings, in which the height of each label is maximized, assuming that the side $s$ is vertical. Polynomial-time algorithms are given for the case of two stacks of labels, and for the case of three stacks of labels under the additional assumption that each $\lopo$-leader that connects a point to a label of the second (third) stack must have its $\lp$-segment in the gap between the first and the second (second and third, respectively) stack of labels. 

\paragraph*{1.5-sided boundary labeling} In this setting, the labels are attached along a single side of the image region $R$, say without loss of generality the right, as in the ordinary $1$-sided boundary labeling. However, an $\lopo$-leader can be routed to the left side of $R$ temporarily and then finally to the right side, that is, its $\lp$-segment is allowed to be on the left side of $R$. Lin et al.~\cite{Lin2009} presented an $O(n^5)$-time algorithm, which -- based on dynamic programming -- computes a length-minimal solution, under the assumption that the position of the labels are fixed and their sizes are uniform; the ports of the labels can be either fixed or sliding. Lin et al.~\cite{LinPTWY11} extended this algorithm to the bend-minimization problem keeping the time complexity unchanged, and they presented an improved algorithm to compute length-minimal solutions in $O(n \log{n})$ time.

\paragraph*{2-sided boundary labeling} In the $2$-sided setting, the algorithmic optimization becomes more complex due to the additional degree of freedom raised by the distribution of the labels to two sides. For uniform labels with either fixed or sliding ports on two opposite sides of the image region $R$, Bekos et al.~\cite{Bekos2007} presented an $O(n^2)$-time algorithm for computing a length-minimal solution. Their approach is based on dynamic programming and exploits the fact that in order to avoid crossings, the vertical order of the sites must be identical to the vertical order of their corresponding labels at each of the two sides of $R$. Under the same set of assumptions, Huang et al.~\cite{Huang2014}, and Fink and Suri~\cite{Fink2016}  proposed less efficient algorithms (which are also based on dynamic programming) to find bend-minimal solutions in $O(n^5)$ time, and length-minimal solutions in the presence of rectangular-shaped obstacles within the image region $R$ in $O(n^{27})$ time, respectively. The time complexity of the algorithm by Fink and Suri~\cite{Fink2016} was improved by Bose et al.~\cite{bckmm-blrd-18} to $O(n^9)$, under the additional assumption that the position of the labels are fixed. Note that if the labels are of non-uniform heights and their positions are not fixed, then both optimization problems mentioned above (i.e., the minimization of the total leader length and the minimization of the total number of leader-bends) become \NP-hard, even in the absence of obstacles~\cite{Bekos2007,Huang2014}. They also remain \NP-hard in the special case in which the input sites are collinear~\cite{Bekos2008c}.

In the many-to-one setting, computing a $2$-sided boundary labeling
with $\lopo$-leaders and minimum number of leader crossings is
\NP-hard, even in the case in which the same number of labels must be
attached on two opposite sides of the image region $R$ and the labels
are of uniform heights with either fixed or sliding
ports~\cite{Lin2008}. For this particular setting, Lin et
al.~\cite{Lin2008} suggested an algorithm with
an approximation factor of at least three that also depends on the
structure of the input instance. If more than one copy of the same
label is allowed, Lin~\cite{Lin2010} provided an $O(n^2)$-time algorithm which
yields a crossing-free routing of the leaders while simultaneously
minimizing the number of used labels.

Finally, in the $2$-sided setting, in which the labels are given along two
adjacent sides of the image region $R$, Bose et
al.~\cite{bckmm-blrd-18} gave an $O(n^{12})$-time algorithm to compute
a length-minimal solution (if any), under the assumption that
the position of the labels are fixed. Notably, their algorithm works
also in the presence of obstacles.

\paragraph*{4-sided boundary labeling}

In the $4$-sided setting, there are significantly fewer results, and they concern boundary labelings of minimal length. To the best of our knowledge, there are no results concerning bend-minimal boundary labelings or boundary labelings that are optimal in terms of some different objective function. 

Bekos et al.~\cite{Bekos2007} presented a  polynomial-time algorithm to compute length-minimal labelings, assuming that the position of the labels around the image region $R$ are fixed (i.e., specified as part of the input). Their algorithm consists of two steps. In the first step, a (not necessarily crossing-free) length-minimal site--label assignment is computed by finding a minimum-cost perfect matching on an appropriately defined bipartite graph (for details refer to Section~\ref{sec:techniques:weighted-matching}). In the second step, the crossings are eliminated such that the total leader length of the assignment computed in the previous step is not affected (i.e., by keeping the solution optimal). The time complexity of the algorithm is dominated by the time needed to compute the minimum-cost perfect matching in the first step and depends on whether the ports of the labels are fixed or sliding. More concretely, in the former case the bipartite graph is geometric, which allows for an efficient minimum-cost perfect matching computation in $O(n^2 \log^3{n})$ time~\cite{Vaidya89}. In the latter case, however, the bipartite graph has no special property, and therefore a minimum-cost perfect matching must be computed using the Hungarian method, which needs $O(n^3)$ time; see, e.g.,~\cite{Kuhn55}. 

We conclude this section by mentioning that the aforementioned algorithm was extended to the case of area features by Bekos et al.~\cite{Bekos2010}. In the case of area features with constant number of corners, the computation of the minimum-cost perfect matching, which determines the complexity of the algorithm, is done on a bipartite graph that is not necessarily geometric, which implies that the Hungarian method must be used (as in the case of sliding label ports above), yielding again a time complexity of $O(n^3)$.

\paragraph{$\ld\lo$- and $\lp\ld$-leaders}
\label{sec:star:do}

Orthodiagonal leaders with one bend have a simple shape, which seems not to introduce clutter in the obtained labelings~\cite{bnn-rlbl-18}. In the framework of boundary labeling, in which the image contour $C_A$ is an axis-aligned rectangle, these types of leaders were introduced and first studied by Benkert et al.~\cite{Benkert2009} back in 2009. On the other hand, if $C_A$ is a circle, then early works that adopt the paradigm of $\ldo$-leaders date back to excentric labeling for interactive labeling of focus regions~\cite{Fekete1999}. 

Let us first assume that $C_A$ is an axis-aligned rectangle.
 As already mentioned, Benkert et al.~\cite{Benkert2009} introduced boundary labelings with $\ldo$-leaders and preliminary observed that a crossing-free solution might not always be feasible. For a general objective function, they proposed a dynamic programming approach, which in $O(n^5)$ time (in $O(n^{14})$ time) determines whether there exist an optimal crossing-free boundary labeling with $\ldo$-leaders, assuming that the ports of the labels are sliding while the positions of the labels are fixed along a single side (along two opposite sides, respectively) of the image region $R$; in the positive case, the labeling can be reported without increasing the time complexity. A significantly improved algorithm was given, when the objective function is the minimization of the total leader length and the labels are along a single side of $R$. For this case, the authors suggested an algorithm, which adopting the sweep-line paradigm determines whether there exists an optimal crossing-free boundary labeling with $\ldo$-leaders in $O(n^2)$ time. 

In order to overcome infeasibility issues, in a subsequent work, Bekos et al.~\cite{Bekos2010b} extended the study of boundary labelings with $\ldo$-leaders by also incorporating $\lod$- and $\lpd$-leaders, and proved that a combination of these types of leaders guarantees the existence of a crossing-free solution. More precisely, they proved that a crossing-free solution (i.e., that is not optimal under some objective function) with $\lod$- and $\lpd$-leader can always be derived by a plane-sweep technique in $O(n \log{n})$ time, assuming that the position of the labels are fixed along the image region $R$. In the same setting, a length-minimal solution can be found in $O(n^3)$ time by appropriately adjusting both steps of the matching-based algorithm given in~\cite{Bekos2007} (see also Section~\ref{sec:star:opo}). 

 Yang et al.~\cite{YangDGM17} observed that while the algorithm by Bekos et al.~\cite{Bekos2007} can ensure the absence of crossings in a length-minimal solution, it cannot guarantee adequate separation between leaders and connection sites or other leaders, which may lead to ambiguity. Hence, they proposed a quadratic program to further improve the output labelings of the algorithm of Bekos~\cite{Bekos2007} and they used them in a framework, called \texttt{MapTrix}, for visualising many-to-many flows between different geographic locations by connecting an OD matrix~\cite{Voorhees2013} with origin and destination maps~\cite{WoodD08}. Löffler et al.~\cite{Loeffler2015} extended the study of boundary labelings with $\ldo$-leaders in the mixed setting, where also internal labels are allowed, and provided a dynamic programming based algorithm to maximize the number of internal labels.

\paragraph{Other polyline leaders}
Few works studied external labelings with polyline leaders that are
neither orthogonal nor orthodiagonal. In text annotation, Kindermann et al.~\cite{Kindermann2014} proposed boundary labelings with $\lo\ls$-leaders, where the first segment of each leader is used to reach the margin of the text, while the second one has a slope such that all labels can be arranged on the margin of the text. Obviously, such labelings can be derived from $\lopo$-labelings by removing the last bend of each leader. Fekete and Plaisant~\cite{Fekete1999}, and Bertini et al.~\cite{Bertini2009}, who studied excentric labeling, also proposed site--label connections with $\lo\ls$-leaders. Here, the first segment of each leader is orthogonal to the boundary of the lens, while the slope of the second segment is chosen such that all labels can be arranged on the boundary of the circular focus region.

\subsection{Curved leaders}\label{sec:star:curved-leaders}
In this section we discuss research that
uses smooth curves as leaders. In the context of automatically
creating metro maps, Wu et al.~\cite{Wu2012} suggested to use 2-sided
boundary labeling and curved leaders based on B-splines to annotate
stations with images. They required that the resulting labeling is
plane, while also minimizing the number of intersections between
leaders and the metro map. At the core of their approach, they used a
general flow-network model to construct such leaders. In a user study
based on eye-tracking, they showed that the curved leaders outperform
straight-line leaders and orthogonal leaders. Wu et al.~\cite{Wu2015}
further extended this approach to 4-sided boundary labeling. To that
end, they first partitioned the image region into four regions using the
method by Bekos et al.~\cite{Bekos2007} and labeled each region
independently with the approach by Wu et al.~\cite{Wu2012}.
Kindermann et al.~\cite{Kindermann2014} used Bézier curves
to connect text annotations with their sites. Their method
first computes a plane straight-line labeling. Afterwards, each leader
is interpreted as a cubic Bézier curve whose control points are
iteratively moved using a force-based method.

\subsection{Empirical evaluation techniques}\label{sec:eval-techniques}
In this section we briefly discuss the state of the art on different
evaluation techniques for empirically examining design decisions
and evaluating labeling techniques. 

The analysis of professional (handmade) drawings, e.g., technical
drawings, visual dictionaries, medical illustrations, is a task that
is neither complicated nor expensive and helps in extracting drawing
criteria for a planned labeling technique.  Niedermann et
al.~\cite{Niedermann2017} extracted drawing criteria in a semi-automatic
way from atlases of human anatomy to generate similarly looking
labelings.  Vollick et al.~\cite{Vollick2007} proposed a system that
automatically learns the drawing style from existing systems.  While
both approaches require that drawings in the specific style already
exist, the analysis of existing drawings can also be used to justify
default drawing criteria satisfied in almost all labeling styles
\cite{Hartmann2004,Ali2005,Hartmann2005,Mogalle2012}.  Moreover,
interviews with domain experts help to justify the extracted criteria
and to make further design
decisions~\cite{Muhler2009,Mogalle2012,Niedermann2017,cb-relgv-18}.

Certainly the most extensive evaluation technique are user studies. In
order to obtain expressive results, an elaborated design, equipment
and sufficiently many participants with different backgrounds are
required. In research on external labeling, user studies are conducted
to assess labeling approaches. Fekete et al.~\cite{Fekete1999}
  investigated the question whether excentric labeling is a reasonable
  alternative for zooming. Čmolík and Bittner~\cite{cb-relgv-18}
  compared their algorithm against label layouts created by
  humans. Madsen et al.~\cite{Madsen2016} presented a study comparing
  different labeling approaches in dynamic settings; to the best of
  our knowledge this is the only investigation that takes multiple
  existing labeling algorithms into account. In contrast, Bertini et
  al.~\cite{Bertini2009}, Mühler and Preim~\cite{Muhler2009}, Pick et
  al.~\cite{Pick2010}, Wu et al.~\cite{Wu2012} and Balata et
  al.~\cite{Balata2014} did not compare their results with other
  approaches, but investigated how participants perform when using
  their approaches and asked them concerning the experiences made when
  using their systems. 
Instead of evaluating a concrete labeling
approach, Barth et al.~\cite{bnn-rlbl-18} conducted a user study to
investigate general properties of external labeling, namely in this
particular case the readability of different leader types.
Depending on the study, the participants were asked to conduct
different types of tasks. As the main purpose of a labeling is to
unambiguously relate features and labels to each other, the most often
proposed task is to associate labels with their features and vise
versa. This is either done by reading tasks~\cite{Bertini2009} or by
selecting particular
labels~\cite{Balata2014,bnn-rlbl-18,cb-relgv-18}. Alternatively, the
participants are asked to check the existence of particular
labels~\cite{Fekete1999} or to solve more complex tasks on the
illustrations that require the use of the labeled
features~\cite{YangDGM17}. Only the user study by Wu et
al.~\cite{Wu2012} used eye-tracking. Typical measures are the response
times and error rates of the participants. Finally, except for the
user study by Čmolík and Bittner~\cite{cb-relgv-18}, all presented
studies take the participants' preferences and personal experiences
with the labeling layouts into account. In particular Pick
et al.~\cite{Pick2010} assessed their approach by conducting an
expert walkthrough in which the participants answered few questions 
after using their~system.

\section{Labeling techniques}\label{sec:techniques}
In this section we discuss algorithmic techniques that are
frequently applied in external labeling. We split them into
non-exact (Section~\ref{sec:techniques:heuristics}) and exact algorithms
(Section~\ref{sec:techniques:exact}). While exact algorithms solve
Problem~\ref{problem:external-labeling} and provide an optimality guarantee for the solution,
non-exact approaches refrain from such guarantees in favor of the running
time. In Section~\ref{sec:techniques:complexity} we discuss complexity
results.

\subsection{Non-exact algorithms}\label{sec:techniques:heuristics}
First we describe algorithmic techniques to solve
Problem~\ref{problem:external-labeling} without necessarily yielding
optimal labelings with respect to a cost function.

\subsubsection{Greedy algorithms}\label{sec:techniques:greedy}
Greedy algorithms are simple local optimization strategies that
iteratively construct solutions by \emph{greedily} extending them by
elements that promise the best improvement of the solution in each
step. For an introduction to greedy algorithms see, e.g,~\cite[Chapter~17]{Cormen2009}.

\paragraph*{Characteristics}
\begin{itemize}
\item Support general multi-criteria cost functions, but typically
  find only local optima.
\item Hard constraints can be easily enforced, but may lead to unlabeled features.
\item Easy to implement and typically fast in practice.
\end{itemize}

\paragraph*{Sketch}
Greedy algorithms in external labeling can be summarized by the
following scheme. For each feature a candidate set of callouts is
parameterized; let $\mathcal C$ denote the union of all those sets.
Starting with an empty labeling $\mathcal L$, the callouts are
\emph{greedily} added to $\mathcal L$ with respect to some given
cost function and hard constraints. More precisely, the
callout~$c\in \mathcal C$ with lowest cost among all callouts in $\mathcal C$ is added
to $\mathcal L$. Afterwards, the callout $c$ and all callouts that
cannot be added to the current labeling $\mathcal L$ without violating
a given hard constraint are removed from $\mathcal C$. The procedure
is repeated until the candidate set $C$ is empty. As result of the
procedure the labeling $\mathcal L$ is returned.  Since in each step
it is guaranteed that $\mathcal L$ is a labeling of the given instance
satisfying all hard constraints, the resulting set $\mathcal L$ is
also a labeling. However, it is not guaranteed that it is the
cost-optimal labeling among all possible labelings. Further, depending
on the hard constraints, in some cases it may happen that not all
features are labeled, even if a complete labeling exists.

\publications{\cite{Hartmann2004},
\cite{Bruckner2005}, \cite{Fuchs2006}, \cite{Gotzelmann2006},
\cite{Lin2008}, \cite{Stein08}, \cite{Cmolik2010},
\cite{wtly-zapalmm-11}, \cite{Mogalle2012}, \cite{Tatzgern2013}, \cite{Heinsohn2014},
\cite{Gemsa2015b}, \cite{cb-relgv-18}}

\subsubsection{Force-based approach}\label{sec:techniques:force-directed}
Force-based approaches are local optimization strategies that have
been originally introduced in the field of graph drawing to construct
layouts of graphs~\cite{e-hgd-84,fr-gdfp-91}. 
According to this approach, 
a graph is modeled as a physical system with forces acting on it, and a good
layout is obtained by an equilibrium state of the system.
In the context of external labeling, the sites, labels,
leaders, and features become the \emph{objects} in the physical system that
interact with each other by forces. The approaches mainly differ in
the objects that are fixed or can freely move, the definition of the
forces, and how an equilibrium is found.

\paragraph*{Characteristics}
\begin{itemize}
\item Supports general multi-criteria cost functions, but typically
  find only local optima.
\item Needs an initial labeling as starting point. 
\item Intuitive formulations as it is based on a mechanic analogy.
\end{itemize}

\paragraph*{Sketch}
A common technique to search for an equilibrium of the system is to
iteratively apply two steps. Firstly, compute for each object~$o$ in the system the
sum $f_o$ of the forces that act on it.  Secondly, move each
object~$o$ along the direction of $f_o$ by an amount that is proportional in the magnitude 
of $f_o$ and a predefined constant $\Delta t$ that may change over time. 
If after a number of iterations the objects do not move any more or a maximum number of
iterations is reached, the system is assumed to be in an equilibrium state. 
We note that the initial configuration of the system, i.e., the one
in the first iteration, strongly influences the quality of the overall
result of the procedure.  For example Ali et al.~\cite{Ali2005} use this iterative
approach as a post-processing step to improve a layout created
beforehand. They introduce forces that optimize the angles between
leaders and the distance between labels and their sites. Similarly as
done in simulated annealing, they start with a large $\Delta t$ and
decrease it with each iteration. Hence, in each iteration the possible
displacement of the labels is reduced. 

Another approach defines potential functions for the objects based on
force fields. The positions of the objects are then chosen such that
they minimize the total potential of the objects. Hartmann et al.~\cite{Hartmann2004} introduce for each label a static
force field such that the attracting force between a label and its
feature increases with increasing distance. Further, a label is
repelled by other objects and the border of the drawing area. The
local minima of the potential function are found by using particles
that move within this force field. Each local minimum is interpreted
as a label candidate of the same feature.  Multiple labels
are placed iteratively. In each iteration a label is placed and fixed
at its preferred position in its force field as described
beforehand. For the remaining labels their force fields are adapted
incorporating the newly placed label as repulsive force.

\publications{\cite{Hartmann2004},\cite{Ali2005},\cite{Hartmann2005},\cite{Pick2010},\cite{Heinsohn2014},\cite{Tatzgern2013},\cite{Balata2014},\cite{Tatzgern2014}}

\subsection{Exact algorithms}\label{sec:techniques:exact}
In this section we describe algorithmic techniques that are used to
solve Problem~\ref{problem:external-labeling} exactly, i.e., with
respect to a given cost function these techniques yield optimal
labelings.

\subsubsection{Dynamic programming}\label{sec:techniques:dynamic-programming}
Dynamic programming is by far the most used technique in the context
of external labeling as evidenced by Tables~\ref{table:sl:properties} and~\ref{table:poly:properties}. The general idea is to find an optimal labeling
using recursion originating from the input instance. In each recursive
step, the optimal labeling of the currently considered instance~$I$ is
composed of optimal labelings obtained from smaller disjoint
sub-instances of $I$. The base case of the recursion is reached when
$I$ does not contain any feature to be labeled and the optimal
labeling of $I$ is trivial. Storing the intermediate solutions in a
table for subsequent look-ups ensures that each sub-solution is
computed only once, which keeps the running time polynomial.  For an
introduction to dynamic programming see, e.g., 
example~\cite[Chapter 16]{Cormen2009}.

\paragraph*{Characteristics}
\begin{itemize}
\item Allows general multi-criteria cost functions.
\item Requires that feasible labelings are crossing-free. 
\item Hard constraints can be easily incorporated.
\item Mostly applied in boundary labeling (see
  Tables~\ref{table:sl:properties} and~\ref{table:poly:properties}).
\item Depending on the actual setting, it tends to yield algorithms that have high asymptotic
  running time and storage consumption.
\end{itemize}

\begin{figure}[t]
  \centering
  \includegraphics[page=7]{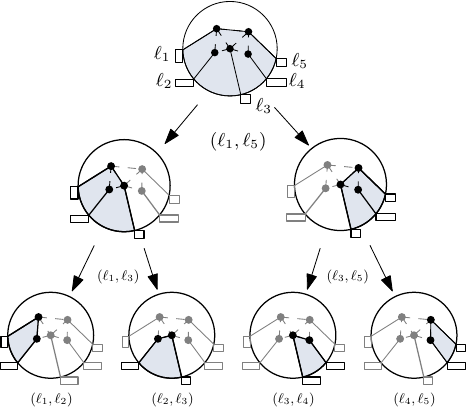}
  \caption{Decomposition tree of an instance for external labeling. The instance (blue) enclosed by the labels $\ell_1$ and $\ell_5$ (top) is decomposed into sub-instances enclosed by the labels $\ell_1$ and $\ell_3$, and $\ell_3$ and $\ell_5$, respectively (middle row). Each of them is again decomposed into two instances, which do not contain any point features anymore. }
  \label{fig:dp:decomposition}
\end{figure}

\paragraph*{Sketch}
In the following we sketch the technique in greater detail providing a
general blue print.  Let $I=(D,F,M)$ be an instance of external
labeling. In the remainder, we call an
instance $I'$ a \emph{sub-instance} of $I$ if the drawing area of
$I'$ is a sub-region of $D$ and the features of $I'$ are a subset of
the features in $F$. Further, two instances are \emph{disjoint} if
their drawing areas do not intersect.

The dynamic programming approach recursively constructs an optimal labeling
$\mathcal L$ of $I$ based on optimal labelings of disjoint
sub-instances of $I$; see Fig.~\ref{fig:dp:decomposition} for an illustration.  To that end, a set $\mathcal I$ of
sub-instances of $I$ is defined (including $I$) such that $I$ can be
recursively decomposed into instances from $\mathcal I$.  More
precisely, two instances $I'$ and $I''$ form a \emph{decomposition} of
$I$ if
\begin{itemize}
\item $I'$ and $I''$ are disjoint sub-instances of $I$,
\item each feature of $I$ is contained either in $I'$ or in $I''$, and
\item both $I'$ and $I''$ contain fewer features than $I$.
\end{itemize}
The recursive decomposition of $I$ is then described as a rooted,
binary tree $T$ over $\mathcal I$ such that
\begin{itemize}
\item each node of $T$ is an instance of $\mathcal I$,
\item each node of $T$ has either no or two children,
\item the children of a node form a decomposition of that node,
\item $I$ is the root of $T$,
\item the leaves of $T$ are empty instances containing no features.
\end{itemize}
We call such a tree a \emph{decomposition tree} of $I$. The actual
choice of $\mathcal I$ strongly depends on the concrete problem setting
and typically requires fundamental insights into the geometric
structure of the problem. In particular, the set $\mathcal I$ is
chosen such that there is a decomposition tree $T$ with
\[
  c_\OPT(I)=\begin{cases}
    c_I, & \text{$I$ is a leaf,}\\
    c_\OPT(I')+c_\OPT(I'')+c_{I',I''} & \text{$I$ has children $I'$ and $I''$,}
            \end{cases}
\]
where $c_\OPT(I)$ denotes the cost of an optimal labeling of $I$, and
$c_I$ and $c_{I',I''}$ are constants.
The existence of such a decomposition tree directly leads to the
following classic dynamic programming approach. We introduce a table
$X$ that contains for each instance $I\in \mathcal I$ an entry $X[I]$
representing the cost of an optimal labeling of $I$, i.e.,
$X[I]=c_\OPT(I)$. Since we consider a minimization problem, we can use
the following recurrence to actually compute $X[I]$.
\[
  X[I]=\begin{cases}
    c_I, & \text{$I$ is empty}\\
    \min_{(I',I'')\in \mathcal D_I}X[I']+X[I'']+c_{I',I''} & \text{otherwise,}
            \end{cases}
\]
where $\mathcal D_I\subseteq \mathcal I \times \mathcal I$ is the
set of all decompositions of $I$ in
$\mathcal I \times \mathcal I$.
We observe that it suffices to consider decompositions from
$\mathcal I \times \mathcal I$, because the decomposition tree only
consists of instances from $\mathcal I$. Further, we compute each
table entry only once -- for example in increasing order with respect
to the number of contained point features. Hence, each table entry
$X[I]$ is computed in $O(|\mathcal I|^2)$ time.
Since there are $|\mathcal I|$ entries in $X$, we obtain
$O(|\mathcal I|^3)$ running time and $O(|\mathcal I|)$ storage
consumption for computing the costs of an optimal labeling of an input
instance.  Using a standard backtracking approach that reconstructs the
solution based on the decomposition tree, we also obtain an optimal
labeling for that instance.
It is crucial to keep $\mathcal I$ as small as possible, because the
running time of the approach strongly depends on the size of
$\mathcal I$.  A common approach is to show that the drawing area of
the instances can be specified by only few parameters over small
domains and to use these parameters to appropriately define the
subinstances.

\begin{figure}[t]
  \centering
  \begin{subfigure}[b]{0.49\linewidth}
    \centering
    \includegraphics[page=3]{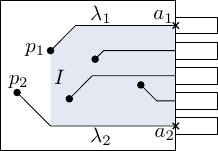}
    \subcaption{\label{fig:dp:do-labeling:instance}}
  \end{subfigure}
  \hfill
  \begin{subfigure}[b]{0.49\linewidth}
    \centering
    \includegraphics[page=4]{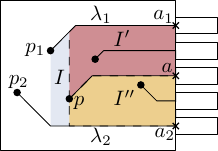}
    \subcaption{\label{fig:dp:do-labeling:decomposition}}
  \end{subfigure}
  \caption{One-sided instance of boundary labeling with
    $\ldo$-leaders. (\subref{fig:dp:do-labeling:instance}) The
    instance $I$ (blue drawing area) is specified by two point
    features $p_1$ and $p_2$, and the two reference points $a_1$ and
    $a_2$. (\subref{fig:dp:do-labeling:decomposition}) The leader
    $\lambda$ with site $p$ and reference point $a$ decomposes the
    instance $I$ into the two disjoint sub-instances $I'$ (red drawing area) and $I''$ (orange drawing area).}
  \label{fig:dp:do-labeling}
\end{figure}

The majority of approaches that are found in literature specify the
drawing area of an instance based on leaders (or on callouts if the
positions of the labels are not fixed). For example, consider 1-sided boundary
labeling for point features and~$\ldo$-leaders such that the label positions
are fixed, the labels touch the right of the boundary rectangle~$R$, and the ports of the labels are also fixed on the midpoint of their left edges;
see Fig.~\ref{fig:dp:do-labeling}. Let $A$ be the set of the induced reference points. We
assume $|A|\geq |F|$, and that the cost function rates single labels.
Similar to Benkert et al.~\cite{Benkert2009} we define the drawing area of a
sub-instance of that problem setting
by a simple polygon specified by two point features $p_1$,
$p_2$ and two reference points $a_1$, $a_2$; see
Fig.~\ref{fig:dp:do-labeling:instance}. More precisely, the
$\ldo$-leader $\lambda_1$ connecting $p_1$ with $a_1$ bounds the area
from above, and the $\ldo$-leader $\lambda_2$ connecting $p_2$ with
$a_2$ bounds the area from below. Further, the area is bounded from
the right by the bounding rectangle and from the left by the vertical
line that goes through the rightmost site of $\lambda_1$ and
$\lambda_2$.  We assume that $\lambda_1$ and $\lambda_2$
neither belong to the drawing area nor to the instance and
observe that all point features of such an instance lie to the right
of the sites of both leaders. Such an instance is uniquely defined by
the choice of the point features and reference points of $\lambda_1$ and
$\lambda_2$.  Thus, all such
instances can be described
by the tuples
$(p_1,a_1,p_2,a_2)\in \mathcal I= F\times A \times F \times A$.
Hence, $\mathcal I$ has size $O(|F|^2|A|^2)=O(|A|^4)$.

Benkert et al.~\cite{Benkert2009} prove that for any input instance there is a
decomposition tree over $\mathcal I$ as defined above. To see that,
consider an optimal labeling $\mathcal L$ of an instance $I$ in
$\mathcal I$; see Fig.~\ref{fig:dp:do-labeling:decomposition}. Let
$\lambda_1$ and $\lambda_2$ be the leaders that bound $I$. We denote
the point features and reference points of $\lambda_1$ and $\lambda_2$ by
$p_1$, $p_2$ and $a_1$, $a_2$, respectively. If $I$ does not contain
any point features, the instance $I$ is a leaf in the decomposition
tree. Further, we have $c_\OPT(I)=0$. Otherwise, if $I$ contains some
point features, let $\lambda$ be the leader with leftmost site in
$\mathcal L$. We denote its site by $p$. The leader splits the
instance into the two disjoint sub-instances $I'=(p_1,a_1,p,a)$ and
$I''=(p,a,p_2,a_2)$. The instances $I'$ and $I''$ are the children of
$I$ in the decomposition tree.  Further, we obtain
$c_\OPT(I)=c_\OPT(I')+c_\OPT(I'')+c(\lambda)$, where $c(\lambda)$
denotes the costs for the leader $\lambda$. Hence, we can apply the
dynamic programming approach as described above.  The set $\mathcal I$
contains $n=O(|A|^4)$ instances and each instance has $m=O(|A|)$
decompositions; recall that we assume $|A|\geq |F|$. Hence, we obtain
$O(|A|^5)$ running time and $O(|A|^4)$ storage consumption. For
$\lpo$-leaders Benkert et al.~\cite{Benkert2009} further show that an
instance can be uniquely specified by two reference points improving the size
of $\mathcal I$ to $n=O(|A|^2)$, which yields $O(|A|^3)$ running time
and $O(|A|^2)$ storage consumption.  Similar dynamic programming
approaches have been considered for other leader-types; see
Tables~\ref{table:sl:properties} and~\ref{table:poly:properties}. Further, while most approaches
specify instances merely based on leaders, few approaches also use
other separators\cite{Kindermann2015,bckmm-blrd-18}.

\publications{\cite{Bekos2006b},\cite{Bekos2007},\cite{Benkert2009},\cite{Lin2009},\cite{Loeffler2010},\cite{LinPTWY11},\cite{Fink2012}, \cite{bckmm-blrd-18},\cite{Huang2014},\cite{Kindermann2014},\cite{Gemsa2015b},\cite{Bekos2015},\cite{Kindermann2015},\cite{Loeffler2015},\cite{Fink2016},\cite{Niedermann2017}}

\subsubsection{Weighted matching}\label{sec:techniques:weighted-matching}

Weighted matching has been applied successfully to instances of external labeling, in which the objective is the minimization of the total leader length, under few reasonable assumptions; see, e.g.,~\cite{Bekos2007,Bekos2010,Bekos2010b}. The idea here is, given an instance of the external labeling problem, to construct an edge-weighted bipartite graph, such that a perfect matching of minimum cost for this graph translates back into a labeling of the original input instance, whose total leader length is minimum. For an introduction to matchings see, e.g.,~\cite[Chapter~12]{Ahuja93}.

\paragraph*{Characteristics}
\begin{itemize}
\item Mostly applies to boundary labeling problems where the objective is to minimize  the total leader length~\cite{Bekos2007,Bekos2010,Bekos2010b}.
\item A post-processing step is usually required to make the computed labelings crossing-free. 
\item The running time of the algorithm is usually dominated by the time needed to solve the matching problem. 
\item Requires uniform labels placed at fixed positions.
\end{itemize}

\paragraph*{Sketch}
We sketch this technique for an instance $I=(D,F,M)$ of the $4$-sided boundary labeling model, in which the feature set $F$ consists of $n$ point features, the leaders are of type-$\ls$, and  both the positions of the labels within the labeling region $A$ and the ports of the labels are fixed~\cite{Bekos2007}.

The main step in this technique is the definition of a weighted bipartite graph $G=(V_F \cup V_L,E,w)$, where $w\colon E\rightarrow \mathbb{R}^+$. The node-set of $G$ is defined as follows. For each point feature $p$ of $F$, graph $G$ contains a node $v(p)$ in $V_F$, and for each label candidate $\ell$, graph $G$ contains a node $v(\ell)$ in $V_L$. The edge-set of $G$ is then defined as the Cartesian product of $V_F$ and $V_L$, that is, $E=V_F \times V_L$. In other words, graph $G$ is complete bipartite. To complete the description of the definition of graph $G$, it remains to describe the weights of its edges. To this end, consider an edge $e$ of $G$. By definition, edge $e$ connects a node $v(p)$ in $V_F$ to a node $v(\ell)$ in $V_L$. The weight $w(e)$ of edge $e$ is defined as the length of the leader $\lambda$ connecting point feature $p$ with the port $\pi(\ell)$ of the label candidate $\ell$. Equivalently, since leader $\lambda$ is of type-$\ls$, the weight $w(e)$ of edge $e$ is set to the Euclidean distance between point feature $p$ and port $\pi(\ell)$.

A labeling $\mathcal{L}$ for instance $I$ that minimizes the total length of the leaders corresponds to finding a minimum-cost perfect matching for $G$, which exists since $G$ is complete bipartite; also, the triangle inequality ensures that no two leaders of $\mathcal{L}$ intersect, i.e., labeling $\mathcal{L}$ is plane. In general, the problem of finding a minimum-cost matching of a weighted bipartite graph can be solved in $O(n^3)$ time using the Hungarian method; see, e.g.,~\cite{Kuhn55}. However, graph~$G$ has a special property that allows this problem to be solved more efficiently, namely, graph $G$ is geometric (that is, drawn on the plane), and the weight of each edge corresponds to the Euclidean distance between its endpoints. For such graphs, the problem of finding a minimum-cost matching can be solved in time $O(n^{2+\delta})$, where $\delta$ can be chosen arbitrarily small~\cite{AES99}.

The technique can also be applied to labels with sliding ports with a minor modification, namely, that the weight of each edge $e=(v(p),v(\ell))$ of $G$ must be the length of the \emph{shortest} leader connecting point feature $p$ with label $\ell$. On the negative side, however, graph $G$ is not geometric anymore, which implies that finding a minimum-cost matching for graph $G$ now takes $O(n^3)$ time. 

We conclude this section by noting that this technique has also been applied to boundary labeling with $\lopo$-leaders~\cite{Bekos2007,Bekos2010}, and with a combination of $\lod$- and $\lpd$-leaders~\cite{Bekos2010b}. However, in these cases a post-processing step is required to eliminate potential crossings between leaders that may appear in the labeling $\mathcal{L}$ computed by the minimum-cost matching for graph~$G$. It is also worth noting that for the case of $\lopo$ -leaders and labels with fixed ports, the minimum-cost bipartite matching problem can be solved more efficiently as observed in~\cite{Bekos2007}, since the bipartite graph~$G$ is geometric with respect to the Manhattan distance, which allows for a solution in $O(n^2 \log^3{n})$ time~\cite{Vaidya89}.

\publications{\cite{Bekos2007}, \cite{Lin2008}, \cite{Bekos2010}, \cite{Bekos2010b}, \cite{Bekos2015}}

\subsubsection{Scheduling}\label{sec:techniques:scheduling}

Boundary labeling has also been related to scheduling problems, in particular to \emph{single-machine scheduling}. Here, a set of $n$ jobs $J_1,J_2,\ldots,J_n$ is given, each of which is to be executed on a single machine. Each job $J_i$ is associated with a processing time $\rho_i$ and a \emph{time window} $(b_i,d_i)$. If a job $J_i$ is processed within its time window, then it incurs no penalty; otherwise, an \emph{earliness} penalty or a \emph{tardiness} penalty  incurs, which is equal to the corresponding deviation. The objective is to determine a schedule, so that either the total earliness-tardiness penalty or the number of penalized jobs is minimized. For an
introduction to scheduling see, e.g.,~\cite[Chapter~1]{Brucker04}.

\paragraph*{Characteristics}
\begin{itemize}
\item Mostly applies to instances of $1$-sided of boundary labeling, in which the objective is to minimize the total leader length or to minimize the total number of leader bends~\cite{Bekos2008c,Gemsa2015b}.
\item The technique has also been used to prove \NP-hardness results~\cite{Bekos2010b,LinPTWY11,Huang2014}. 
\item The main idea is that the length of each label corresponds to the processing time of a corresponding job, while the objective to be optimized is expressed as a sum of the earliness and of the tardiness penalties of the jobs.
\end{itemize}

\paragraph*{Sketch}
We sketch this technique on an instance  $I=(D,F,M)$ of boundary labeling with type-$\lopo$ leaders and sliding labels of non-uniform sizes that must be placed on one side on the boundary of the image region $R$, say the right side. For each $i=1,2,\ldots,n$, point feature $p_i$ of $I$ is associated with a job $J_i$, whose processing time is $h_i$ and whose  due window is $(y_i-h_i, y_i+h_i)$, where $y_i$ and $h_i$ denote the $y$-coordinate of $p_i$ and the height of the label of $p_i$, respectively. 

Assume first that $S^*$ is a schedule that minimizes the number of penalized jobs and consider the labeling $\mathcal{L}$ obtained by setting the $x$-coordinate of the lower left corner of the label of the point feature $p_i$ to the starting time of job $J_i$ in $S^*$. Then, it is not difficult to see that $\mathcal{L}$ has the minimum number of leader bends, because for each $i=1,2,\ldots,n$ the leader connecting point feature $p_i$ to its label is of type $o$ (that is, without any bend) if and only if job $J_i$ is completely processed in its time window in $S^*$. On the other hand, if $S^*$ minimizes the total earliness-tartiness penalties, then a symmetric argument implies that labeling $\mathcal{L}$ is optimal in terms of the total length of the leaders. These simple linear-time reductions imply that finding $1$-sided boundary labelings with sliding labels of non-uniform sizes that are either optimal in terms of the total number of leader bends or in terms of the total leader length can be found in $O(n^2)$ and $O(n \log{n})$ time, respectively~\cite{Bekos2008c,KOULAMAS1996190}.

We finally note that the above reduction has also been used to prove \NP-hardness mostly for non-uniform labels. The main idea here is that the order of the labels (and as a result the order of the jobs to be executed on the machine) may be variable when the leaders are not of type $opo$; see, e.g.,~\cite{Bekos2010b,LinPTWY11,Huang2014}. Hence, finding a schedule that minimizes the total earliness-tardiness penalty or the number of penalized jobs becomes \NP-hard, which gives rise to corresponding \NP-hardness reductions.

\publications{\cite{Bekos2008c}, \cite{Bekos2010b}, \cite{LinPTWY11}, \cite{Huang2014}, \cite{Gemsa2015b}}

\subsubsection{Plane sweep}\label{sec:techniques:sweep}

Plane sweep is a well-known technique with several applications, mainly in computational geometry. This technique has been proven to be useful also in boundary labeling since several instances of this problem have purely geometric flavour (in particular, for uniform labels with position and fixed ports); see \slSweepLabel in Table~\ref{table:poly:properties}. We stress, however, that the technique mostly applies when the goal is to find a crossing-free solution efficiently (that is, without taking into account any optimization criterion). A notable exception is the $O(n\log{n})$-time algorithm by Benkert et al.~\cite{Benkert2009} to compute 1-sided length-minimal boundary labelings with $\lp\lo$-leaders.  For a
an introduction to plane-sweep approaches see, e.g.,~\cite{BergCKO08}.

\paragraph*{Characteristics}
\begin{itemize}
\item Mostly applies to variants of boundary labeling with uniform labels, fixed reference points and ports.
\item Usually results in crossing-free boundary labelings. 
\item The complexity of the obtained algorithms is usually $O(n \log{n})$, since a sorting of the input point features (e.g., from top to bottom or from left to right) is required.
\end{itemize}

\paragraph*{Sketch}
We describe this technique on a simple variant of 1-sided boundary labeling with $\ls$-leaders, in which the external labels are of uniform size with fixed reference points and ports aligned on a single, say right, side of the image region $R$.  The algorithm is incremental, as it processes the labels from bottom to top in order to compute a labeling in which no two leaders cross~\cite{Bekos2007}. 

More precisely, let $\pi_1,\pi_2,\ldots,\pi_n$ be the ports of the labels as they appear from bottom to top. Then, for $i=1,2,\ldots n$, the $i$-th port $\pi_i$ is connected via an $\ls$-leader to the first unlabeled point feature that is hit by a ray $r_i$ that emanates from $\pi_i$, initially points vertically downwards, and is rotated around $\pi_i$ in clockwise order. Suppose that a crossing occurs between any two leaders. But then the rotating ray would have found the corresponding point features in the reverse order, which contradicts the fact that the two leaders cross. 
A straightforward implementation yields a time complexity of $O(n^2)$, which can be improved to $O(n \log{n})$ using standard computational geometry techniques.

We conclude the description of this technique by mentioning that the algorithm presented above can be extended to compute corresponding 4-sided boundary labelings, under the same set of restrictions~\cite{Bekos2007}. The idea is to partition the image region $R$ into a particular number of convex non-overlapping sub-regions (using again appropriate rotating lines), such that the number of point features in the interior of each sub-region equals the number of labels on its boundary, which implies that the $1$-sided algorithm can be applied to each of the computed sub-regions independently.

\publications{\cite{Bekos2007}, \cite{Benkert2009}, \cite{Bekos2010b},  \cite{Bekos2010b}, \cite{Bekos2011}, \cite{Lin2010}, \cite{Noellenburg2010}, \cite{LinPTWY11}, \cite{Fink2012}, \cite{Kindermann2014}}

\subsubsection{Mathematical programming}\label{sec:techniques:math-programming}
Mathematical programming is a
general global optimization approach. A mathematical program consists
of a set of variables, a set of inequalities and an optimization
function. The goal is to find an assignment of the variables such that
all inequalities are satisfied and the value of the optimization
function is (depending on the application) minimized or
maximized. Often, these approaches make use of integer or binary
variables. While this particularly provides the possibility of easily
formulating combinatorial problems such as
Problem~\ref{problem:external-labeling}, integer variables, generally
speaking, turn the optimization problem into an \NP-hard one. Still,
powerful solvers for mathematical programs have been developed, which
can be used to solve a broad variety of instances in reasonable
time. For an introduction to mathematical programming see, e.g.,~\cite{Vanderbei04}.

\paragraph*{Characteristics}
\begin{itemize}
\item General technique that can be used to formulate many (\NP-hard)
  optimization problems.
\item Solvers are available, but their running times are
  hard to predict.
\item Common technique to obtain optimal solutions for the purpose of
  evaluating optimality gaps of heuristics for the same problem.
\end{itemize}

\paragraph*{Sketch}
We briefly sketch a basic variant of a mathematical program for
Problem~\ref{problem:external-labeling}. More precisely, we present an
integer linear programming formulation (ILP), i.e., a mathematical
program that consists of integer variables, linear inequalities and a
linear objective function. Alternatively, one may also use quadratic
programming formulations. Instead of linear cost-functions they also
allow the formulations of quadratic cost-functions. While this
supports the formalization of more complex problems, this usually also
comes with a significantly higher running time.

We assume that we are given a set $F$ of
features and for each feature $f\in F$ a set $L_f$ of callout
candidates; we denote the union of all those candidate sets
by~$L$. Further, let $c\colon L \to \mathbb R$ be a cost function that
rates each candidate in $L$. We aim at a labeling
$\mathcal L \subseteq L$ such that for each feature $f\in F$ exactly
one callout $\gamma\in L_f$ is contained in $\mathcal L$ and
$\sum_{\gamma \in \mathcal L} c(\gamma)$ is minimized among all those
labelings.

Based on the work by Barth et al.~\cite{bnn-rlbl-18}, we use the
following ILP formulation to express this optimization problem. We
note that this is a standard formulation to express assignment
problems. For each candidate $\gamma \in L$ we introduce a binary
variable $x_{\gamma}\in \{0,1\}$. We interpret the variable~$x_\gamma$
such that $x_{\gamma}=1$ if the callout is selected for the labeling
and $x_\gamma=0$ otherwise. We further introduce the following linear
constraints.
\begin{align}
  x_{\gamma} + x_{\gamma'} \leq 1 & \text{ for each $\gamma,\gamma'\in L$ that exclude each other}\\
  \sum_{\gamma \in L_f}\gamma=1 &\text{ for each feature $f\in F$}
\end{align}
Constraint~(1) ensures that if two callouts exclude each other due to
some given hard constraint (e.g., because they intersect), at most one
of them is added to the labeling. Further, Constraint (2) enforces
that for each feature exactly one candidate is selected. Subject to
Constraints~(1) and~(2) we minimize the total costs
$\sum_{\gamma \in L} c(\gamma)\cdot x_{\gamma}$.
The resulting labeling is then defined as
$\mathcal L=\{\gamma \in L \mid x_\gamma =1\}$. We note that Barth et
al.~\cite{bnn-rlbl-18} use a similar formulation to minimize the total
leader length in boundary labeling.

\publications{\cite{Bekos2011},\cite{Gemsa2015b},\cite{YangDGM17},\cite{bnn-rlbl-18}}

\subsection{Complexity results}\label{sec:techniques:complexity}

In contrast to internal labeling, where the majority of the problems turn out to be \NP-hard, several external labeling problems admit polynomial-time solutions, as we have seen so far. In external labeling the problems become usually \NP-hard, when the labels are non-uniform or their positions are not fixed. Next, we present some of these negative results grouped by the problem used to prove hardness. For an
introduction to NP-hardness see, e.g.,~\cite{GareyJ79}.

The well-known, weakly \NP-complete \textsc{Partition} problem~\cite{GareyJ79} has been used in several reductions in boundary labeling. Bekos et al.~\cite{Bekos2007} observed that the task of assigning non-uniform labels to two opposite sides of the image region $R$, say to two vertical sides, is in one-to-one correspondence with the \textsc{Partition} problem, if the label heights sum up to twice the height of $R$. In the context of panorama labelings, Gemsa et al.~\cite{Gemsa2015b} employ a reduction from the \textsc{Partition} problem (that uses the same principle ideas as an earlier one of Garrido et al.~\cite{GarridoIMPRW01}) to prove that determining whether a panorama labeling with vertical leaders exists is \NP-hard. A conceptually similar reduction is given by Bekos et al.~\cite{Bekos2008c} in the context of labeling collinear sites with non-uniform labels, but directly from the labeling problem by Garrido et al.~\cite{GarridoIMPRW01}. In the context of boundary labeling with obstacles, Fink and Suri~\cite{Fink2016} suggest another reduction from the \textsc{Partition} problem to prove that determining whether a $1$-sided labeling with $\lpo$-leaders and non-uniform labels exists is \NP-hard, even in the case of a single obstacle. Finally, a slightly more involved reduction from a variant of the \textsc{Partition} problem has been suggested by Bekos et al.~\cite{Bekos2006b} to prove that the multi-stack boundary labeling problem with non-uniform labels is \NP-hard.

Bekos et al.~\cite{Bekos2010b}, Lin et al.~\cite{LinPTWY11} and Huang et al.~\cite{Huang2014} proved that different variants of the boundary labeling problem, in which the labels are non-uniform and the objective is the minimization of the total leader length, are \NP-hard by employing conceptually similar reductions from an \NP-complete scheduling problem commonly known as \textsc{Total Discrepancy}~\cite{GareyTW88}. According to this problem, the jobs have different processing times but they all share a common \emph{midtime}, in which the first half of each job must be completed; the goal is to compute a schedule that minimizes the total earliness-tardiness penalty (see also Section~\ref{sec:techniques:scheduling}).

In the many-to-one boundary labeling setting, Lin et al.~\cite{Lin2008} prove that the problem of minimizing the number of crossings between leaders can be expressed as a specific instance of the \NP-hard \textsc{Two-Layer Crossing Minimization} problem~\cite{EadesW94}, according to which the goal is to compute a two-layer drawing of a given bipartite graph with minimum number of crossings. Bekos et al.~\cite{Bekos2015} proposed a reduction from the \textsc{Fixed-Linear Crossing Number} problem~\cite{MasudaNKF90} to prove that the problem of minimizing the number of crossings between leaders remains \NP-hard, even in the case where several leaders to the same label can share a common (horizontal) segment. Recall that the fixed-linear crossing number problem asks for a linear embedding of a given graph, in which the order of the vertices is fixed along a line $L$ and the edges must be drawn either completely above or completely below $L$ such that the number of edge-crossings is minimum. 

\publications{\cite{Bekos2006b}, \cite{Bekos2007}, \cite{Lin2008}, \cite{Bekos2008c}, \cite{Bekos2010b}, \cite{LinPTWY11}, \cite{Huang2014}, \cite{Bekos2015}, \cite{Fink2016}}

\section{Discussion and conclusion}\label{sec:discussion}
The previous two sections provided an in-depth discussion of the state of the art on external labeling algorithms. Our literature review identified that two mostly independent research communities have studied external labeling problems extensively: the visual computing community as well as the algorithms community. 
In visual computing the
vast majority of research considers straight-line leaders and presents
heuristics taking a large set of drawing criteria into account. In
these practical works the concrete application and deployment
of the approaches play a major role. In contrast, the algorithms
community focuses on proving formal quality guarantees in labeling models limited to fewer drawing criteria. Algorithmic and geometric properties and results are primary, while abstraction makes the concrete
application become secondary. In these algorithmic works the results are formal models and exact algorithms, mainly for polyline leaders and boundary labeling.
This raises the question what the two communities can learn from
each other and how research efforts can be joined to create synergies. In the following we discuss this question in greater detail and state ten open challenges in external labeling. In these challenges we have not included immediate follow-up problems of individual papers (e.g, improve an algorithm's complexity or extend a method to a broader setting). Rather, these challenges emerge from the discussion that we have developed so far. They concern questions that are missing in the existing literature and, according to our expertise, need attention.

As evident from Section~\ref{sec:star}, a plethora of different labeling approaches have been developed and
evaluated in the visual computing community. Still, direct comparisons between different results are are missing from the literature, as opposed, e.g., to traditional map labeling, where such comparisons are available both in terms of algorithmic efficiency, but also in terms of the quality of the produced labelings~\cite{DBLP:journals/tog/ChristensenMS95}. 
As a result, the approaches are
mostly assessed by discussing labelings of selected example illustrations and
3D models. Hence, even if the approaches consider similar settings
(possibly in different applications), one can hardly assess which of
the approaches is better in the sense that it yields the more readable
or more aesthetically pleasing labeling. Since the evaluations that have been conducted so far are mainly preliminary (see Section~\ref{sec:eval-techniques}), we deem the systematic and quantitative evaluation of the existing approaches to be one of the most important challenges. In order to obtain sustainable results, this particularly requires to make the results comparable. Especially practitioners will benefit from empirically founded readability results and guidelines when choosing a labeling style and an algorithm for their labeling tasks.

\begin{op}\em 
  Systematically evaluate and compare different approaches for external labeling.
\end{op}

This is where the theoretical works on boundary labeling have their
particular strength. They use a common formal structure that is
divided into two parts: the labeling model (including all
parameters, hard and soft constraints) and the algorithms computing
labelings within this particular model. Hence, the formalism
admits simple comparisons of the models' features. Further, due to the
clear distinction between model and algorithm, an evaluation can be split into two independent parts, i.e., the assessment of the labeling
model and the assessment of the labeling algorithm. The unified
taxonomy presented in Section~\ref{sec:model} is the first step
towards generalizing this structure from boundary labeling to more
general external labeling models, enabling a better comparison of different approaches.

Multiple labeling models have been proposed in both communities.
As observed in Sections~\ref{sec:star:sl} and~\ref{sec:star:poly}, in the algorithms community these models are
mostly specified by defining a particular leader type and a precise but limited objective function (e.g., the total leader length) that takes into account general drawing principles (e.g., proximity, planarity, and straightness/low detour~\cite{p-wagehu-97,wpcm-cmga-02}). On the other hand, in the visual computing community user studies, experts interviews and the manual analysis of handmade drawings are common tools to extract and assess various additional
drawing criteria, which are combined and balanced using multi-criteria optimization heuristics. As evident from Section~\ref{sec:eval-techniques}, in both communities an extensive and systematic
comparison of the labeling models is missing. 
Currently, criteria are often selected by common sense, experience, and following general cognitive and visualization principles and guidelines~\cite{r-tsi-17}. While this is reasonable, we believe that
establishing a set of quantitative, well-justified quality
measures specific to external labeling would be of great help when comparing and choosing between different labeling models. It lends itself to include the already extracted drawing criteria (e.g., short, crossing-free leaders) as a basis for such measures, but extend this also towards more global measures such as balance, harmony, and interplay between image and callouts. Introducing
new measures requires a systematic justification and ranking.

\begin{op}\em 
  Establish and rank quantitative quality and aesthetics measures for evaluating and comparing external labelings.
\end{op}

Assuming suitable and computable quality measures are established, we
can compare different labeling models by computing and evaluating
labelings for large sets of input instances. For example, one might
aim at labelings that exclusively optimize a single quality measure
such that two different labeling models can be compared with respect
to this measure. We note that this procedure is independent from the
actual labeling algorithm, but only takes the labeling model into
account. However, this raises several problems. First of all, we need
algorithms that compute mathematically optimal solutions for labeling
problems. While for boundary labeling a large variety of such
algorithms exist (as evident from Section~\ref{sec:star:poly},
  and Tables~\ref{table:sl:properties}
  and~\ref{table:poly:properties}), for contour labeling only the
algorithm by Niedermann et al.~\cite{Niedermann2017} is known. For the
general external labeling problem there exist a large number
  of heuristics that perform well in specific application scenarios,
  but there is a lack of exact algorithms as evident from
  Section~\ref{sec:star:sl}. On the other hand, exact algorithms
based on established techniques such as dynamic programming or
mathematical programming are expected to have high running times and
thus limited practical applicability. Hence, the development of
alternative algorithmic techniques that perform well in practice is
needed.  From a theoretical point of view it is also interesting to
find lower bounds on the time complexity for both the underlying
computational problems and the developed algorithms.

\begin{op}\em 
  Develop efficient algorithms for optimally solving external labeling problems  that are applicable in practice.
\end{op}

We deem such algorithms to be helpful both for evaluating labeling models and heuristics, and for producing high-quality labelings used in
professional books such as atlases of human anatomy, where somewhat longer computation times can be tolerated.  On the other
hand, for dynamic settings such as those that appear in the visualization of 3D models the computation of labelings in real-time is
required. Further, complex constraints that improve temporal coherence
must be satisfied. For such applications fast heuristics seem to be the first
choice. We stress here that the vast majority of the existing heuristics (see Section~\ref{sec:star:sl}) depend highly on the application domain and as a result they are tailored to the specifications of the problem, which they are designed to solve (e.g., \cite{Fuchs2006,Mogalle2012,Balata2014}). Therefore, it is of importance to develop more generic heuristics of broader scope that can be deployed in several applications.

\begin{op} 
\em  Develop more generic, less problem-specific heuristics and algorithms for external labeling. 
\end{op}

Dynamic and interactive settings have been extensively considered by
the visual computing community, while the algorithms community has mainly
considered static labelings. A promising research direction is to
build up on that knowledge to integrate temporal coherence and user
interaction into formal models and to develop algorithms with formal
quality guarantees for dynamic settings. The absence of such algorithms is greatly evident from Tables~\ref{table:sl:properties} and~\ref{table:poly:properties}; in Table~\ref{table:sl:properties} observe that the intersection of S4 (temporal coherence) and S7.2 (algorithm type: exact) is empty, while in Table~\ref{table:poly:properties} there is only one entry (i.e.,~\cite{Noellenburg2010}) in row P9.1 (algorithm type: exact) for dynamic external labelings, which focuses on a very restricted case as discussed in Section~\ref{sec:star:poly}.

\begin{op}\em 
  Develop efficient exact algorithms for computing dynamic, temporally coherent, and interactive external labelings.
\end{op}

To obtain comparable and reproducible experimental results in labeling research, it is necessary to establish benchmark sets (which are common in several other fields, e.g., traditional map labeling~\cite{WW97}, graph clustering~\cite{BaderMS0KW14}, route planning~\cite{BlumS18} and graph drawing~\cite{BattistaD13}) and make them available for experiments. Depending on the problem setting, these benchmark sets may consist of geographic data (as, e.g., those in~\cite{WW97}), illustrations, maps, or 3D models each annotated with feature sets of varying type, density, and complexity. The need of these benchmarks is even more evident from the fact that several papers in the literature contain sample labelings obtained by implementations of different approaches (refer to~S6.1 in Table~\ref{table:sl:properties} and to~P8.2 in Table~\ref{table:sl:properties}) but no systematic/quantitative evaluations.

\begin{op}\em 
  Establish general and representative benchmark sets for external labeling.
\end{op}

Establishing benchmarks will further facilitate the development
and the systematic evaluation of (possibly much faster) heuristics, as
their performance will be easy to be evaluated by simply comparing the
optimal labelings and the results of these heuristics with respect to
different quality measures. 
To simplify the process of such
evaluations we suggest to develop a common external labeling framework and interface that allows to easily plug in different models and algorithms. This particularly may motivate researchers to make their source code available to the scientific community (e.g., only two of the 41 implementations reported in Tables~\ref{table:sl:properties} and~\ref{table:poly:properties} are publicly available in source code~\cite{Fekete1999,Bruckner2005} and just few others provide JavaScript implementations or legacy Java applets).
We note that the open source library \emph{OpenLL} is currently being developed and provides an API specification for label rendering in 2D and 3D graphics~\cite{Limberger2018}. As the library has its focus on rendering, only simple placement algorithms are provided so far. We deem this library to be one possible starting point for tackling Challenge~7.  

\begin{op}\em 
  Develop a generic framework for models and algorithms on
  external labeling.
\end{op}

We pointed out that there is a division between strong practical work
using straight-line leaders and a significant amount of theoretical
work on polyline leaders. This can be partially explained by the fact
that straight-line leaders are the shortest and simplest visual
association between sites and labels, while polyline leaders possess
complaisant geometric properties that can be exploited for exact
algorithms. Nevertheless, it is far from clear which leader type is to
be preferred. In a first formal user study on leader readability,
Barth et al.~\cite{bnn-rlbl-18} showed that $\lpo$-leaders outperform
$\ls$-leaders and that $\lopo$-leaders lag far behind. However, those
experiments were conducted for boundary labeling using blank
background images. Hence, in other applications, such as text
annotations, $\lopo$-leaders seem to be a better choice because they
can run between the text lines, while $\lpo$- or $\ls$-leaders pass
through the text.  Furthermore, $\ls$-, $\lp\lo$-, $\ld\lo$- and
$\lp\ld$-leaders, as well as curved leaders are all frequently found
in professional information graphics. They arguably differ in their
aesthetic appeal, which is not only influenced by their geometry, but
also by the type and style of the background image.  The few
  existing guidelines as described in Section~\ref{sec:visual-aspects}
  are valuable resources when assessing different options of external
  labelings, but they do not cover all types of labelings and are
  primarily based on practical
  experience~\cite{w-digtpssm-00,r-tsi-17}. There is a lack of formal
  empirical studies and readability results to confirm or adapt these
  guidelines.

\begin{op}\em 
Establish evidence-based guidelines and rules to determine which leader type suits which application and task best.
\end{op}

Finally, we consider the combination of internal and external labels
to be a fruitful research field, as comparatively little work has been
done on this topic. From the algorithmic perspective only few special cases of mixed labeling have been considered so far that mainly combine very simple internal and external labeling models~\cite{Bekos2011,Loeffler2010,Loeffler2015}. 
From a practical, visual perspective the question
that arises is when to use or not to use external rather than internal labels. In the same direction, it is also worth studying more sophisticated mixed labeling models, which allow combinations of internal and external labels, but also support \emph{internal} labels connected to their features by short leaders; see, e.g.,~\cite{Bell2001,LuboschikSC08}. 

\begin{op}\em 
  Develop rules to address the dichotomy between internal and external labels, i.e., when to use which type of labels.
\end{op}

\begin{op}\em 
  Broaden the research on mixed labeling by combining more sophisticated labeling models for internal and external labels.
\end{op}

In summary, both in the visual computing community and in the
algorithms community a multitude of specializations and variants of
external labeling have been considered from two different
perspectives. This survey gave a comprehensive overview over both
fields discussing the research and algorithmic techniques that have been presented over the last two decades. At the same time it shall pave the way to join both research directions creating new synergies and pushing forward the progress on external labeling.

\paragraph*{Acknowledgments} We thank the anonymous reviewers for their helpful comments and suggestions to improve this survey.  

\bibliographystyle{eg-alpha}
\bibliography{strings,references}

\end{document}